\documentclass[a4paper,fleqn,usenatbib]{mnras}
\usepackage[utf8]{inputenc}
\usepackage[T1]{fontenc}
\usepackage{ae,aecompl}

\usepackage{color}
\usepackage{hyperref}
\usepackage{amssymb}
\hypersetup{colorlinks,citecolor=blue}
\usepackage{xcolor}
\usepackage{graphics,graphicx}
\usepackage{epsf} 
\usepackage{amsmath,amssymb}
\usepackage{longtable}
\usepackage{deluxetable}
\def\arcsec{$^{\prime\prime}$}
\def\~{$\sim$}

\def\Lq{\textquotedblleft}

\newcommand{\myemail}{yogesh.chandola@gmail.com}
\title[H{\sc i} absorption towards radio AGNs]{ H{\sc i} absorption towards low luminosity radio-loud AGNs of different accretion modes and \textit{WISE} colours}
\author[Chandola Y.\& Saikia D.J.]{Yogesh Chandola$^{1}$
\thanks{\myemail} and
 D.J. Saikia$^{2,3}$ \\
$^{1}$ National Astronomical Observatories, Chinese Academy of Sciences, Beijing 100012, China\\
$^{2}$ Cotton College State University, Panbazar, Guwahati, 781 001, India\\
$^{3}$ NCRA, TIFR, Post Bag 3,  Ganeshkhind, Pune 411 007, India}

\date{\today}
\pubyear{2016}
 
\begin{document}
\label{firstpage}
\pagerange{\pageref{firstpage}--\pageref{lastpage}}
\maketitle
\begin{abstract}
	H{\sc i} absorption studies of active galaxies enable us to probe their circumnuclear regions and the general interstellar medium, and study the supply of gas which may trigger the nuclear activity. In this paper, we investigate the detection rate of H{\sc i} absorption on the nature of radio galaxies based on their emission-line spectra, nature of the host galaxies based on the \textit{WISE} colours and their radio structure, which may help understand the different accretion modes. 
We find significant difference in distributions of W2$-$W3 colour for sources with  H{\sc i} absorption detections and non-detections. We report a high detection rate of H{\sc i} absorption in the  galaxies with \textit{WISE} infrared colours W2$-$W3 $>$ 2, which is typical of gas-rich systems, along with  a compact radio structure.  The H{\sc i} detection rate for low-excitation radio galaxies (LERGs) with W2$-$W3 $>$ 2 and compact radio structure is high (70.6$\pm$20.4 \%). In HERGs, compact radio structure in the nuclear or circumnuclear region could give rise to absorption by gas in the dusty torus in addition to gas in the interstellar medium. However, higher specific star formation rate (sSFR) for the LERGs with W2$-$W3 $>$ 2 suggests that H{\sc i} absorption may be largely due to star-forming gas in their hosts.
 LERGs with extended radio structure tend to have significantly lower values of W2$-$W3 compared to those with compact
structure. Extended radio sources and those with W2$-$W3 $<$ 2 have low H{\sc i} detection rates.
\end{abstract}
\section{Introduction} 
    
 Optical spectroscopic studies of  radio galaxies \citep{1979MNRAS.188..111H,1994ASPC...54..201L, 2010A&A...509A...6B,2012MNRAS.421.1569B}  led to the division of radio-loud AGNs (Active Galactic Nuclei) into two categories, HERGs (High Excitation Radio Galaxies) and LERGs (Low Excitation Radio Galaxies). This dichotomy is believed to be due to differences in the accretion modes of HERGs and LERGs \citep{2010A&A...509A...6B, 2012MNRAS.421.1569B, 2012ApJ...757..140S}.  In the high-excitation or quasar mode or radiatively efficient mode of accretion, Eddington ratio is greater than 1\%, while it is less than 1\% in the low-excitation mode or radio mode or radiatively inefficient mode of accretion \citep{2014ARA&A..52..589H}.  In HERGs, accretion in the nuclear regions takes place through geometrically thin, optically thick accretion disks \citep{1973A&A....24..337S,1973blho.conf..343N} while for LERGs, this thin disk is absent  and hence the radiatively inefficient accretion mode is believed to operate \citep{1994ApJ...428L..13N, 1995ApJ...452..710N, 2014ARA&A..52..529Y}. HERGs or sources with higher accretion rate are  galaxies with  younger stellar populations, growing central black hole masses  and higher rate of central star formation \citep{2008MNRAS.384..953K,2012MNRAS.421.1569B}. 
Contrary to this, LERGs or sources with lower excitation and lower accretion rate are  galaxies of higher stellar masses, higher  black hole masses and with redder optical colours, consisting older stellar population than HERGs \citep{2008MNRAS.384..953K,2012MNRAS.421.1569B}.
 While LERGs contain both FR\,I and FR\,II type sources \citep{1974MNRAS.167P..31F}, HERGs are predominantly FR\,II \citep{2012MNRAS.421.1569B, 2014ARA&A..52..589H}. 
Differences are also seen in their luminosities. A majority of LERGs have a radio luminosity less than $\sim$ 10$^{26}$ W Hz$^{-1}$ at 1.4 GHz, while HERGs dominate above this luminosity
\citep{2012MNRAS.421.1569B}. 
     
 It has been suggested that different kinds of  fuelling processes may be dominant in different kinds of AGN \citep{2004IAUS..222..235M}.
 While interactions and major mergers are thought to be behind the fuelling of HERGs \citep{2014MNRAS.445L..51T, 2015ApJ...806..147C}, for low luminosity radio-loud AGNs, which would include most of the LERGs, accretion of hot halo ISM or IGM gas or minor mergers have been suggested as possible fuelling mechanisms by a number of authors \citep{2006MNRAS.372...21A, 2007MNRAS.376.1849H, 2007NewAR..51..168B, 2008A&A...486..119B, 2015MNRAS.451L..35E}. However, although many theoretical models are based on the framework where HERGs are fuelled by major mergers, the observational evidence is mixed, and different processes may be operating \citep{2014ARA&A..52..589H}. In the case of LERGs, major mergers are unlikely to be the triggering and fuelling mechanism \citep{2015MNRAS.451L..35E}. Recent studies \citep{2014MNRAS.444.3408Y} of early-type galaxies find presence of cold atomic/molecular gas in $\sim$40 \% of galaxies in the nearby Universe, and there is also evidence that early-type galaxies with dust lanes have higher chances to host emission-line AGNs \citep{2012MNRAS.423...59S}. \cite{2015MNRAS.449.3503D} argue that early-type galaxies with dust lanes have acquired their dust and gas from recent external minor mergers. For sources in clusters of galaxies, accretion of hot
halo gas which has cooled also provides a viable mechanism \citep{2014ARA&A..52..589H}. In their infra-red study of a sample of southern 2 Jy  radio galaxies, \cite{2014MNRAS.445L..51T} found host galaxies to have enough cold ISM fuel needed for the central AGN, but they argue that only having enough fuel is not sufficient for triggering the AGN activity, and it also depends on the kinematics and distribution of cold gas.   

H{\sc i} absorption studies towards radio-loud AGNs have been a prominent tool to study  cold neutral hydrogen gas in the central regions of the host galaxies of these sources \citep{1989AJ.....97..708V, 2001MNRAS.323..331M, 2003A&A...404..871P, 2003A&A...404..861V, 2006MNRAS.373..972G,2010MNRAS.406..987E, 2011MNRAS.418.1787C, 2012MNRAS.423.2601A, 2013MNRAS.429.2380C, 2015A&A...575A..44G}. Studying the H{\sc i} properties of these sources with different radio luminosities, optical and infrared properties may provide clues towards further understanding and distinguishing between the accretion modes. However, most of the H{\sc i} absorption studies have been predominantly towards higher radio luminosity sources (L$_\mathrm{1.4 {GHz}}$ $>$ 10$^{26}$ WHz$^{-1}$), due to senstivity limitations of the instruments while observing flux density limited samples, except in a few cases e.g. \citep{2010MNRAS.406..987E,2011MNRAS.418.1787C, 2014A&A...569A..35G, 2015A&A...575A..44G}, where luminosities have been lower than 10$^{26}$ WHz$^{-1}$. In our studies \citep{2011MNRAS.418.1787C}, we observed the Compact Radio sources at Low Redshift (CORALZ) core sample of 18 sources, which was compiled by \cite{2004MNRAS.348..227S}, with the sources having flux densities larger than 100 mJy at 1400 MHz and angular sizes less than 2 arcsec. With a small sample and large statistical uncertainties, H{\sc i} properties were found to be similar to those in higher luminosity sources. A recent study by \cite{2014A&A...569A..35G, 2015A&A...575A..44G} of a much larger sample of 101 sources has H{\sc i} absorption detected towards 32 sources.    

In this paper, we explore possible dependences of H{\sc i} absorption properties of the sources on HERGs/LERGs and hence on the accretion mode, source size, and nature of the host galaxy as reflected by the infrared colours. We consider the sample observed uniformly by  \cite{2015A&A...575A..44G} and the classification into high-excitation and low-excitation radio galaxies done by \cite{2012MNRAS.421.1569B} for a large sample of sources using the SDSS optical spectroscopic data (DR7; \citealt{2009ApJS..182..543A}) and the \emph{Faint Images of the Radio Sky at Twenty Centimeters survey} (FIRST; \citealt{1995ApJ...450..559B}) and \emph{NRAO VLA Sky Survey} (NVSS; \citealt{1998AJ....115.1693C}) radio data.
We also used the\emph{ Wide-field Infrared Survey Explorer} (\emph{\textit{WISE}}; \citealt{2010AJ....140.1868W})  infra-red (IR) archival data to look for relation between IR  and H{\sc i} gas properties of LERGs and HERGs. \textit{WISE} data provide mid-infrared photometry at different wavelengths (W1: 3.4 $\mu$m, W2: 4.6 $\mu$m, W3: 12 $\mu$m, W4: 22 $\mu$m). \textit{WISE} colour plots are useful to distinguish between optical AGN activity (W1$-$W2; \citealt{2013MNRAS.436.3451S}) and emission from warm dust heated due to AGN or star formation (W2$-$W3). W2$-$W3 colour can be also used as an indicator of star formation history in a galaxy \citep{2012ApJ...748...80D}, and hence helpful in distinguishing IR \lq early' type and IR \lq late'  type galaxies \citep{2010AJ....140.1868W}. We explore how the H{\sc i} absorption gas properties change with W2$-$W3 colour for LERGs and HERGs.    
\begin{figure*}
  \centering
    \hbox{   
    \includegraphics[scale=0.44]{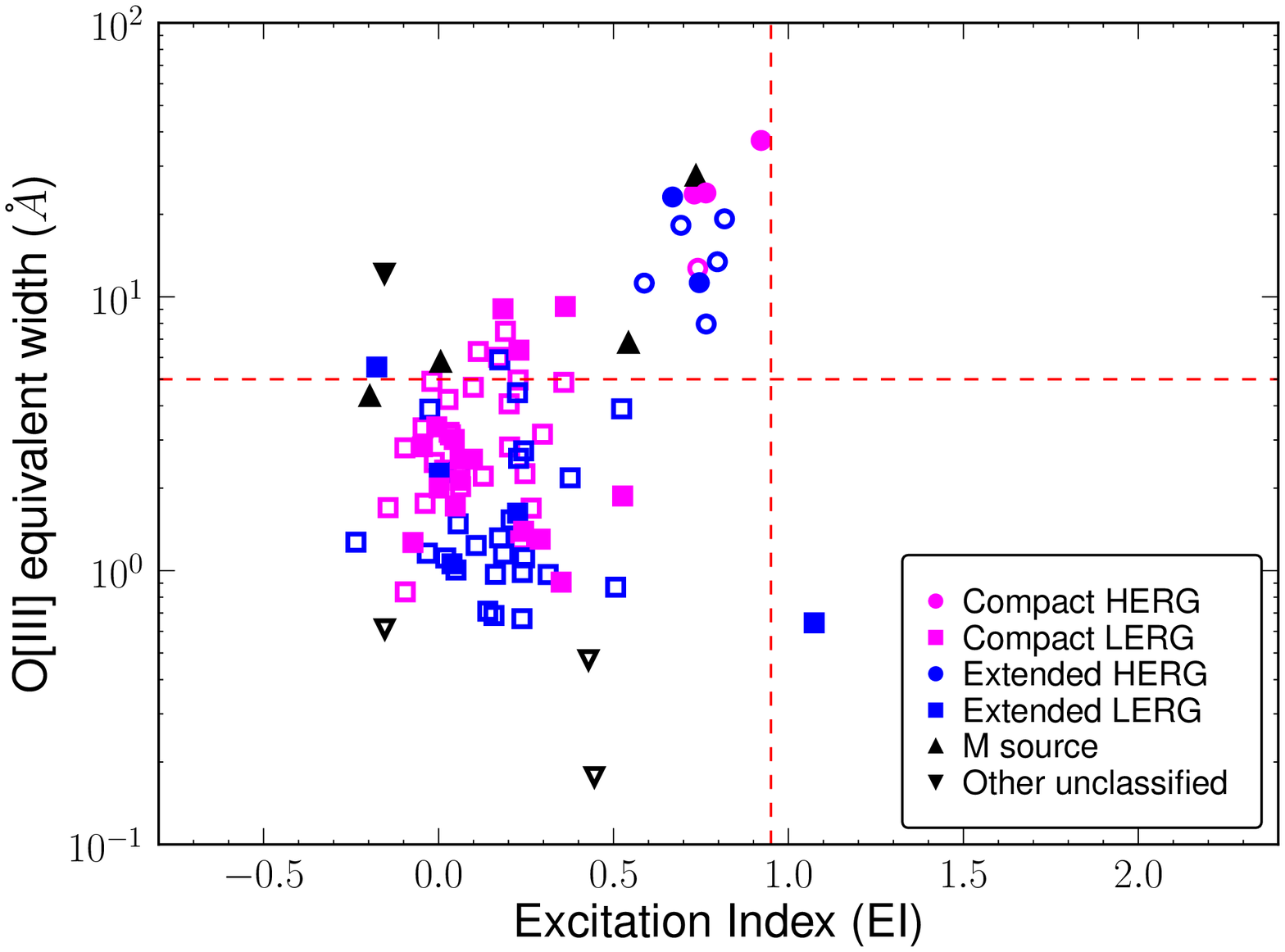}
    \includegraphics[scale=0.44]{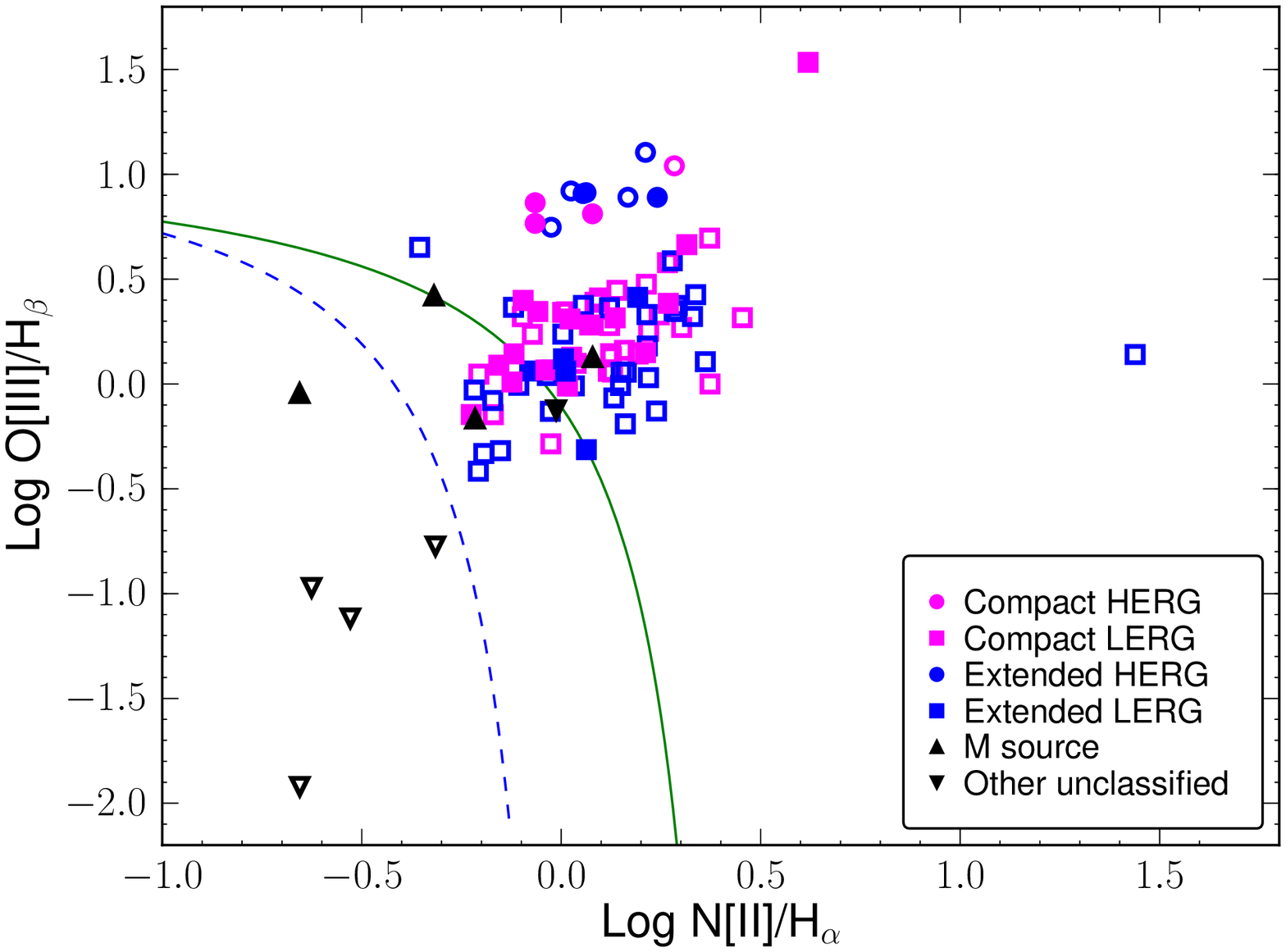}  
     }
   \caption{
Left: O[{\sc iii}] equivalent width vs. excitation index (for 91 sources with all 6 emission lines) with filled symbols (detections) and empty symbols(non-detections), vertical line represents EI=0.95 while horizontal line is for O[{\sc iii}] equivalent width = 5 \AA; Right:  Log O[{\sc iii}]/H$\beta$ vs. Log N[{\sc ii}]/H$\alpha$ (for 99 sources with atleast 4 emission lines), solid green curve is the
\protect\cite{2001ApJ...556..121K,2006MNRAS.372..961K} dividing line between AGNs and composite (SF+AGN) galaxies while dashed blue curve is the \protect\cite{2003MNRAS.341...33K} dividing line between star-forming and composite galaxies.   
    }
     \label{fig1}
  \end{figure*}  
     
\begin{figure*}
    \centering
    \hbox{
     \includegraphics[scale=0.44]{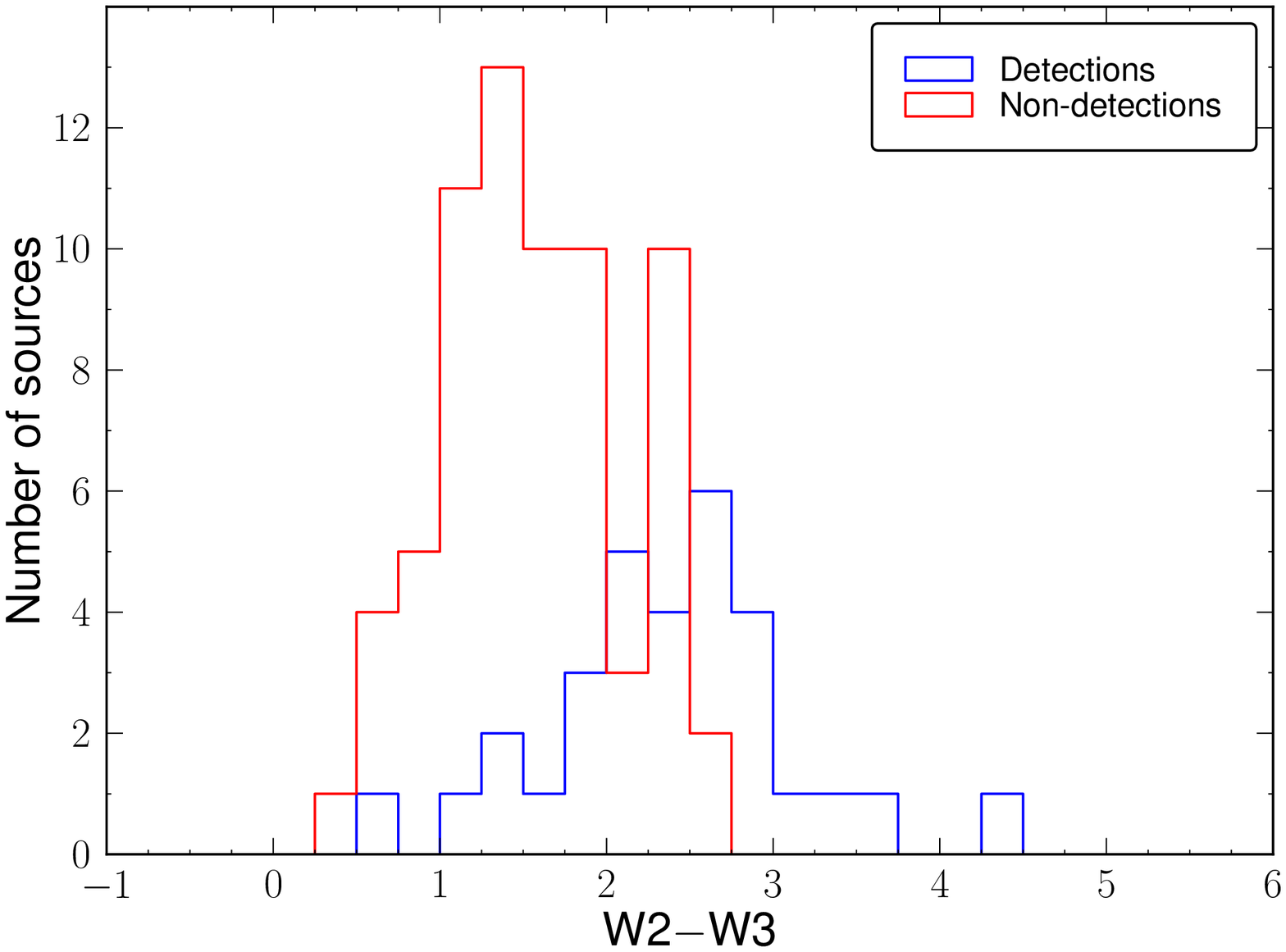}
     \includegraphics[scale=0.44]{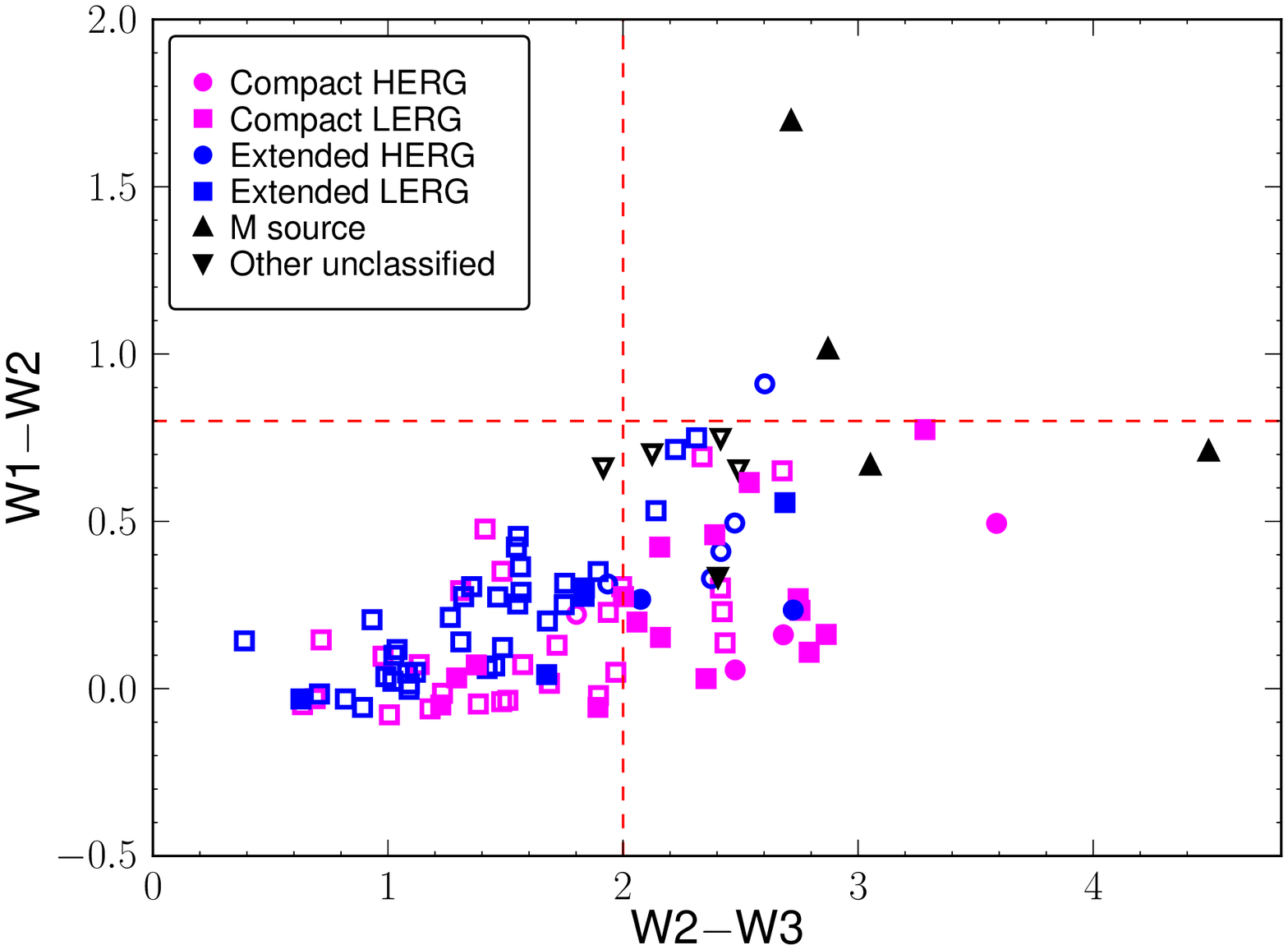}
     }
    \hbox{
    \includegraphics[scale=0.44]{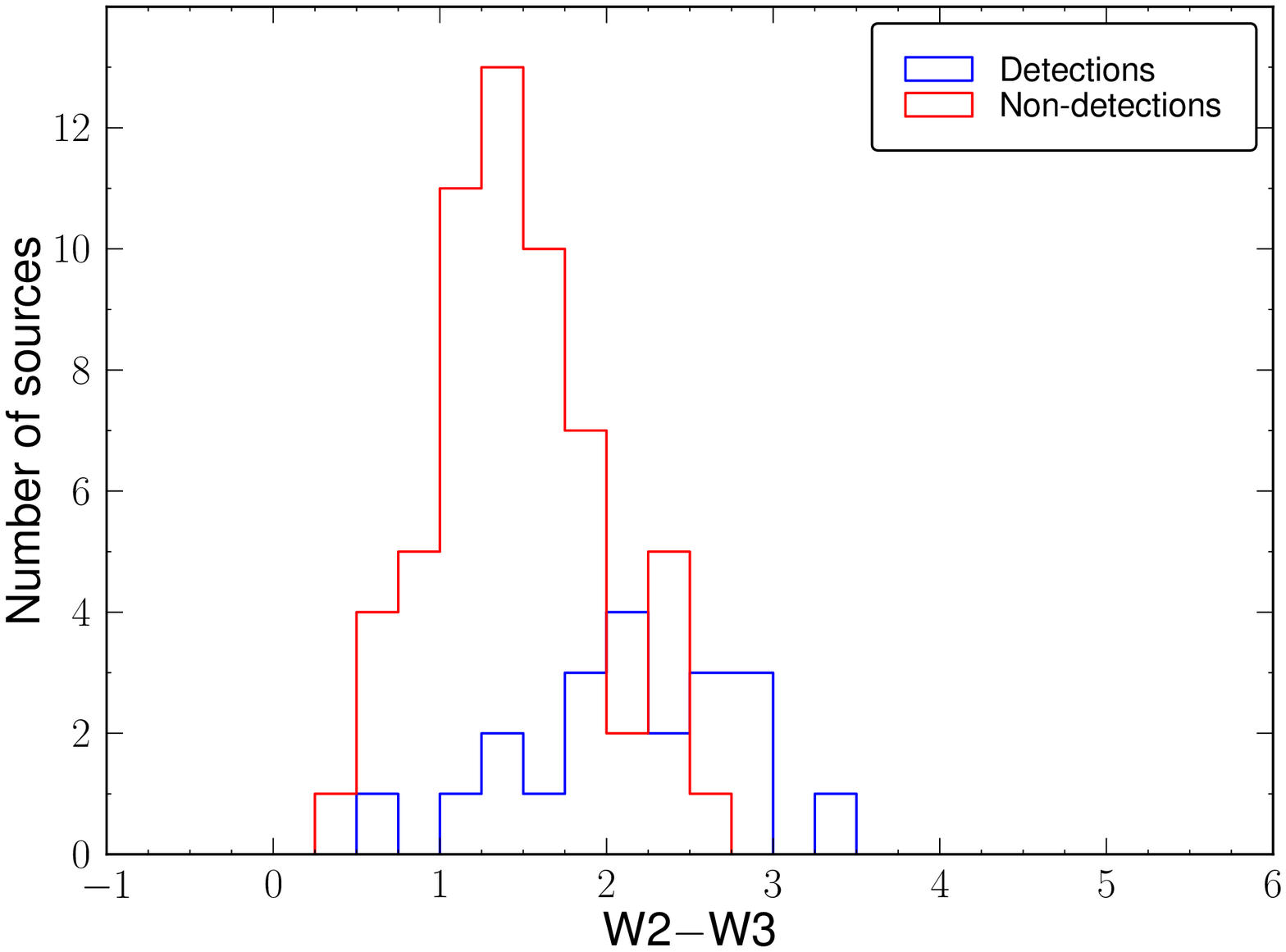}
     \includegraphics[scale=0.44]{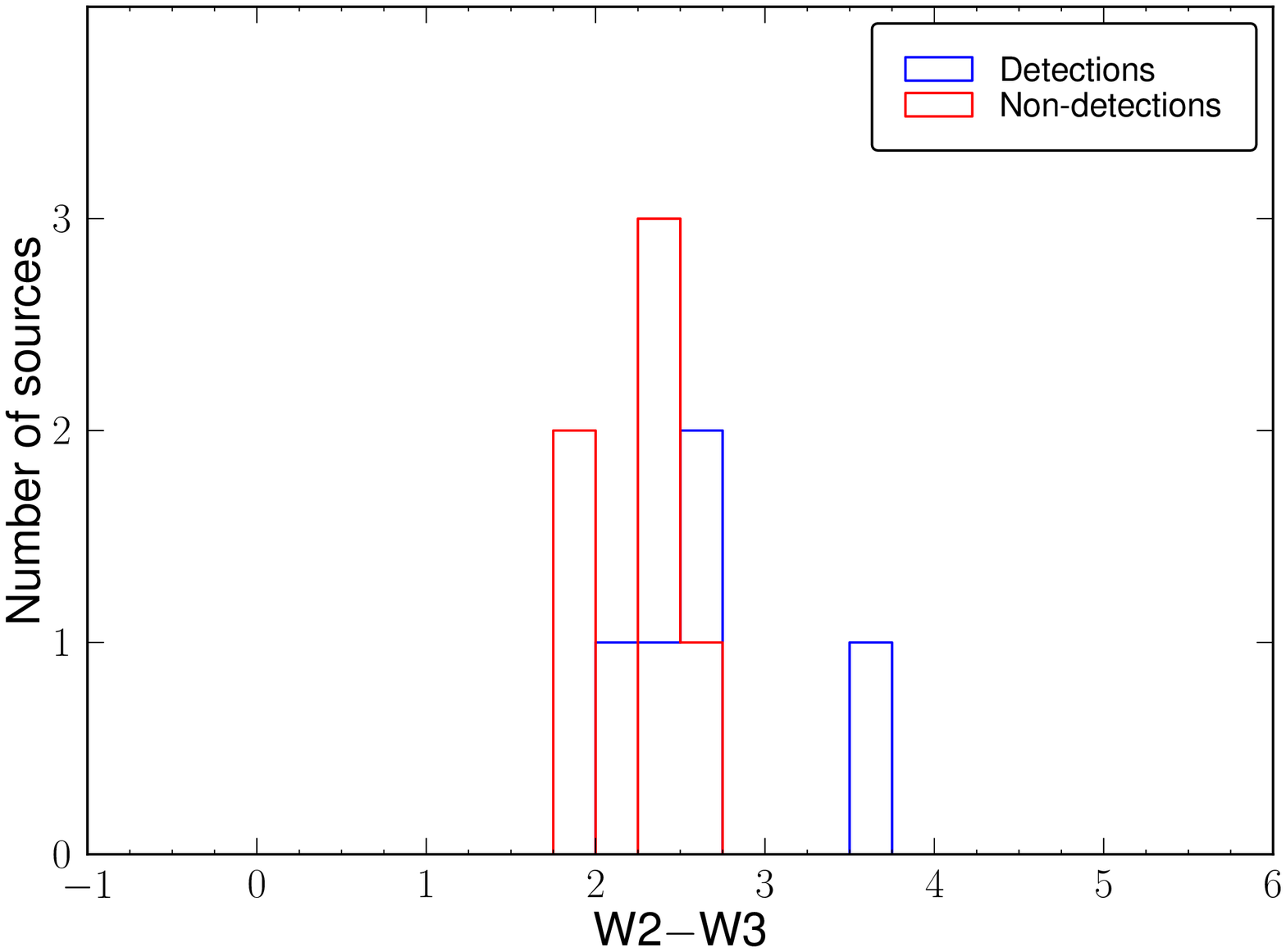}
     }    
    \caption{
Top Left: Distributions of W2$-$W3 colours for H{\sc i} absorption detections (blue) and non-detections (red); 
             Top Right: W1$-$W2 vs. W2$-$W3 plot for detections (filled symbols) and non-detections (empty symbols) for the different categories of objects. Horizontal red dashed line shows W1$-$W2 $=$ 0.8 and vertical red dashed line shows W2$-$W3 $=$ 2   
 Bottom:    Distributions of W2$-$W3 colours for H{\sc i} absorption detections and non-detections shown separately for LERGs (left) and HERGs (right).                          
    }
     \label{fig2}
  \end{figure*}  
 
\section{Sample selection} 
Using the Westerbork Synthesis Radio Telescope (WSRT), \cite{2014A&A...569A..35G, 2015A&A...575A..44G} studied H{\sc i} in absorption towards a sample of 101 radio sources, selected by cross matching SDSS \citep{2000AJ....120.1579Y}, and FIRST-NVSS data, and considering sources in the red-shift range 0.02 $<$ z $<$ 0.23 and FIRST flux density greater than 50 mJy. The radio luminosity of these sources at 1.4 GHz are in the range 10$^{23-26}$ WHz$^{-1}$, due to the lower red-shift and lower flux density level. \cite{2014A&A...569A..35G} observed each source for $\sim$4 hours and have reported that the sources with H{\sc i} detections and non-detections have statistically similar flux density distributions. Therefore their optical depth sensitivities are also similar, and hence non-detections are unlikely to be due to a poor signal to noise ratio, but due to a lower optical depth compared with the detections.

The optical spectroscopic classification of almost all these sources were done using the work of  
\cite{2012MNRAS.421.1569B} who constructed a sample of 18286 radio sources (15300 are radio-loud AGNs) by cross-matching SDSS optical spectroscopic data (DR 7) and FIRST-NVSS radio data. They used the \Lq excitation index", EI $= \rm{log}_{10}(\rm{[OIII]}/H\beta) -
\frac{1}{3}\left[\rm{log}_{10}({\rm [NII]} / H\alpha) +
  \rm{log}_{10}(\rm{[SII]} / H\alpha) + \rm{log}_{10}(\rm{[OI]} /
  H\alpha)\right]$, defined by \cite{2010A&A...509A...6B}  to classify sources as LERGs (EI $<$ 0.95-1$\sigma$) and HERGs (EI $>$ 0.95+1$\sigma$) where all 6 emission lines were detected.   
    For other cases, they used multiple approaches such as using \cite{2006MNRAS.372..961K}, \cite{2010MNRAS.403.1036C} diagnostics, and equivalent width of O[{\sc iii}] emission lines (those with absolute values greater than 5\AA ~were classified as HERGs).  We cross matched \cite{2015A&A...575A..44G} sample with \cite{2012MNRAS.421.1569B} sample and find 100 sources to be common.  We used \textsc{topcat} software package \citep{2005ASPC..347...29T} for this purpose. Of these, 11 are classified as HERGs, 81 as LERGs and 8  
are unclassified. Again, of the 100 sources, we find that of the 31 sources detected in H{\sc i}, 5 are HERGs, 21 LERGs, 5 unclassified, while of the 69 sources without any H{\sc i} detection, 6 are HERGs, 60 LERGs and 3 unclassified. 

 Further cross-matching these 100 sources with the \textit{WISE} catalogue \footnote{\url{http://vizier.cfa.harvard.edu/viz-bin/Cat?II/328}}\citep{2013yCat.2328....0C, 2013wise.rept....1C} using search radius of 10\arcsec, we have mid-IR data on all 100 sources. \textit{WISE} observations have been more sensitive at 3.4 $\mu$m and 4.6 $\mu$m than at higher wavelengths of 12 $\mu$m and 22 $\mu$m \citep{2013AJ....145...55Y}. Hence, we have more reliable magnitudes for W1 and W2 than W3 and W4. In our sample for six sources (J075607+383401, J080042+321728, J083411+580321, J110305+191702, J122823+162613, J130621+434751), we have 2$\sigma$ upper limits on W3 magnitudes. However for completeness we have used these upper limits along with the measured values ($>$3$\sigma$) of W3 magnitudes to estimate W2$-$W3 colours. Our results remain unchanged even if these 6 are excluded. 
  
Since nearly all sources have their optical colours g$-$r $>$ 0.7 \citep{2015A&A...575A..44G}, the host galaxies are optically of early type except a few unclassified sources and one H{\sc i} non-detection, J094542+575748, classified as LERG. Of the  8 unclassified  sources, 5 sources are with H{\sc i } detection. Of these 5 sources, 4  have been marked as M sources (associated with mergers or have blue optical colours) by \cite{2015A&A...575A..44G}, which include two ongoing mergers (UGC05101 and IC883 or UGC 8387), one star forming galaxy (blue colour, g$-$r $=$ 0.38 and from SDSS spectra) B2 1229+33  and one with Seyfert like morphology (IRAS 13384+4503). Fifth unclassified H{\sc i} detection, J160338+155402 is a compact radio source associated with red elliptical galaxy, identified as FSRQ (Flat Spectrum Radio Quasar) by \cite{2007ApJ...671.1355T}. Of the three unclassified non-detections, two sources, J081827.29+281403 and J122121.9+301037, marked as compact radio sources by \cite{2015A&A...575A..44G} have been identified as QSOs (Quasi Stellar Objects) from their SDSS spectra. Another unclassified non-detection, a compact radio source, J120303+603119, is associated with an E/S0 galaxy. 
 
 To be sure about LERG/HERG classification by \cite{2012MNRAS.421.1569B}, we checked online\footnote{\url{http://wwwmpa.mpa-garching.mpg.de/SDSS/DR7/}} optical emission line catalogue of SDSS spectral line data by the Max Planck Institute for Astrophysics, and the Johns Hopkins University (hereafter MPA-JHU; \citealt{2004astro.ph..6220B}). We have provided tables with these measurements for our sample in online version of the paper as  supplementary material (see sample Tables 1, 2 and 3 in Appendix). 
Of the 100, 91 sources have all 6 emission lines detected, used for calculating excitation index. 
Of the remaining 9, 5 sources,  J083139+223423, J091218+483045, J122122+301037, J130622+434751, J154912+304716, have O[{\sc i}] equivalent width (EQW) $>$ 0 \AA \hspace{1mm} indicating absorption. Of these 5,  J122122+301037 has been categorized as unclassified and remaining 4 classified as LERGs by \cite{2012MNRAS.421.1569B}. Two sources, J080042.0+321728 and J110305.8+191702, both LERGs, have S[{\sc ii}] equivalent width $>$ 0 \AA\hspace{1mm} indicating absorption. J082028+485347, classified as LERG, has S[{\sc ii}] and O[{\sc i}] absorption lines. The ninth source, J151641.6+291810, classified as LERG, has reliable measurements available only for H$\beta$, O[{\sc i}] and O[{\sc iii}] emission lines, where O[{\sc iii}] equivalent width is 1.049$\pm$0.175 {\AA} and line flux is 26.04$\pm$6.53 $\times$10$^{-17}$ erg s$^{-1}$cm$^{-2}$. It is an extended radio source (radio structural classification discussed in next section) and LERGs are known to have weak emission lines, sometimes difficult to detect, we continue with its classification as LERG. We further plotted O[{\sc iii}] Equivalent width vs. Excitation Index for 91 sources  with all six emission lines and  Log O[{\sc iii}]/H$\beta$ vs. Log N[{\sc{ii}}]/H$\alpha$ \citep[BPT diagram]{1981PASP...93....5B} for 99 sources with atleast 4 emission lines, and checked the location of different sources in these diagrams (Fig.~\ref{fig1}). We find in the equivalent width vs. excitation index plot, most of the HERGs lie close to the dividing line of EI $=$ 0.95 but with O[{\sc{iii}}] equivalent width $>$ 5 \AA,  and clear division between two classes in BPT diagram.  Except a few possible outliers classified as LERGs, e.g. J094542+575746, J142210+210554, J154912.3+304716, J155611.6+281133, the classification between LERGs and HERGs is consistent in both plots. The sources, J154912.3+304716 and J155611.6+281133, are extended radio sources. So we continue with them classified as LERGs. J142210+210554 is a compact radio source and have low flux measurements on all 6 lines, specially H$\beta$  measurements are low with high errors but O[III] equivalent width is $<$ 5 \AA. So we continue with its classification as LERG. However, we exclude the source J094542+575748, classified as LERG, and put it in unclassified category along with the 8 sources already without any classification from \cite{2012MNRAS.421.1569B}, because it is located in star-forming region in BPT diagram and also have bluer colour. This gives us  26 sources detected in H{\sc i}, 5  HERGs and 21 LERGs,  and 65 sources without any H{\sc i} detection, 6 HERGs and 59 LERGs, for our analysis. However, we will also show the positions of all 9 unclassified sources in the plots. 
 
\section{Results }
 \begin{figure*}
    \centering
     \hbox{
    \includegraphics[scale=0.44]{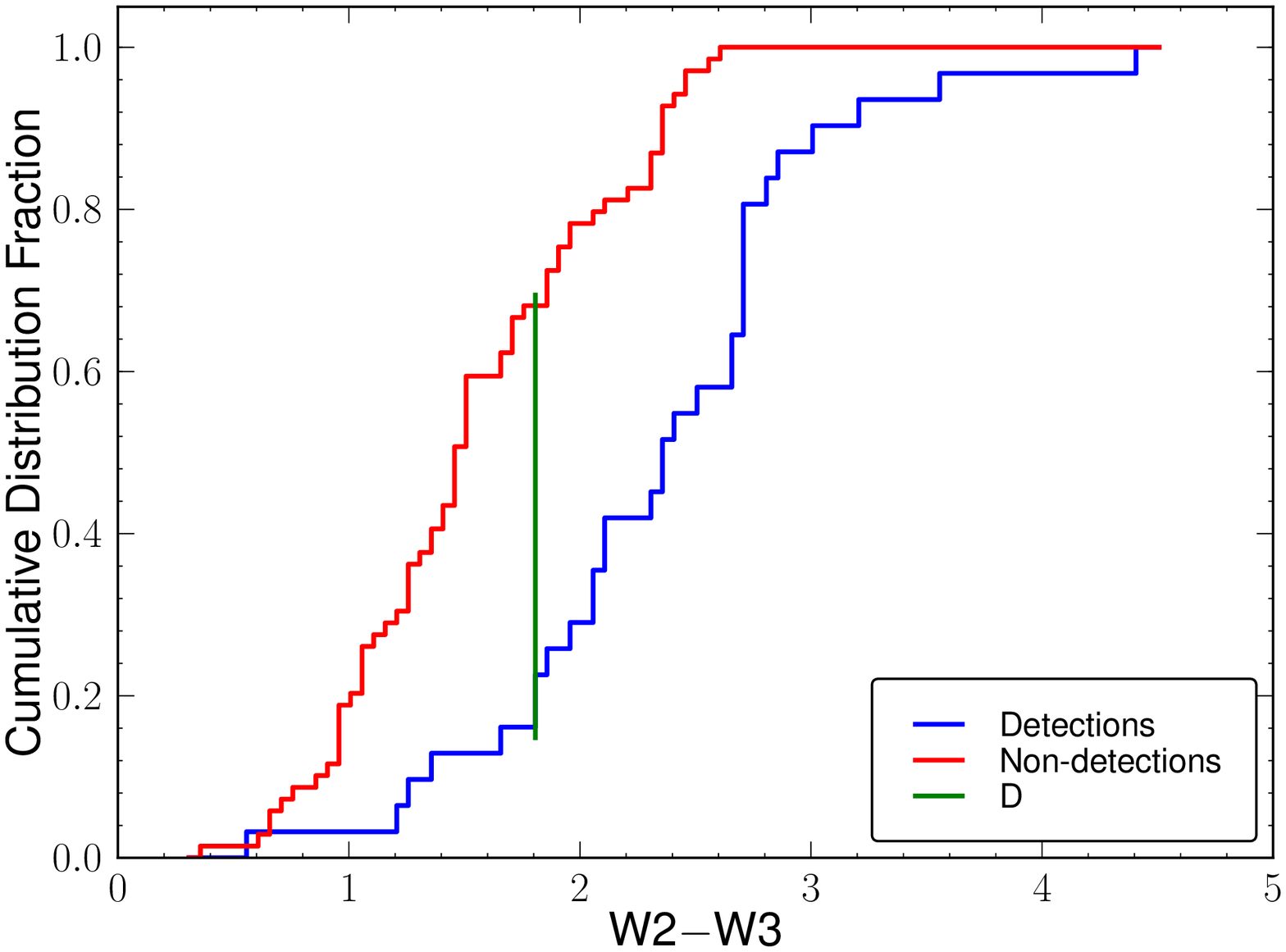}
      \includegraphics[scale=0.44]{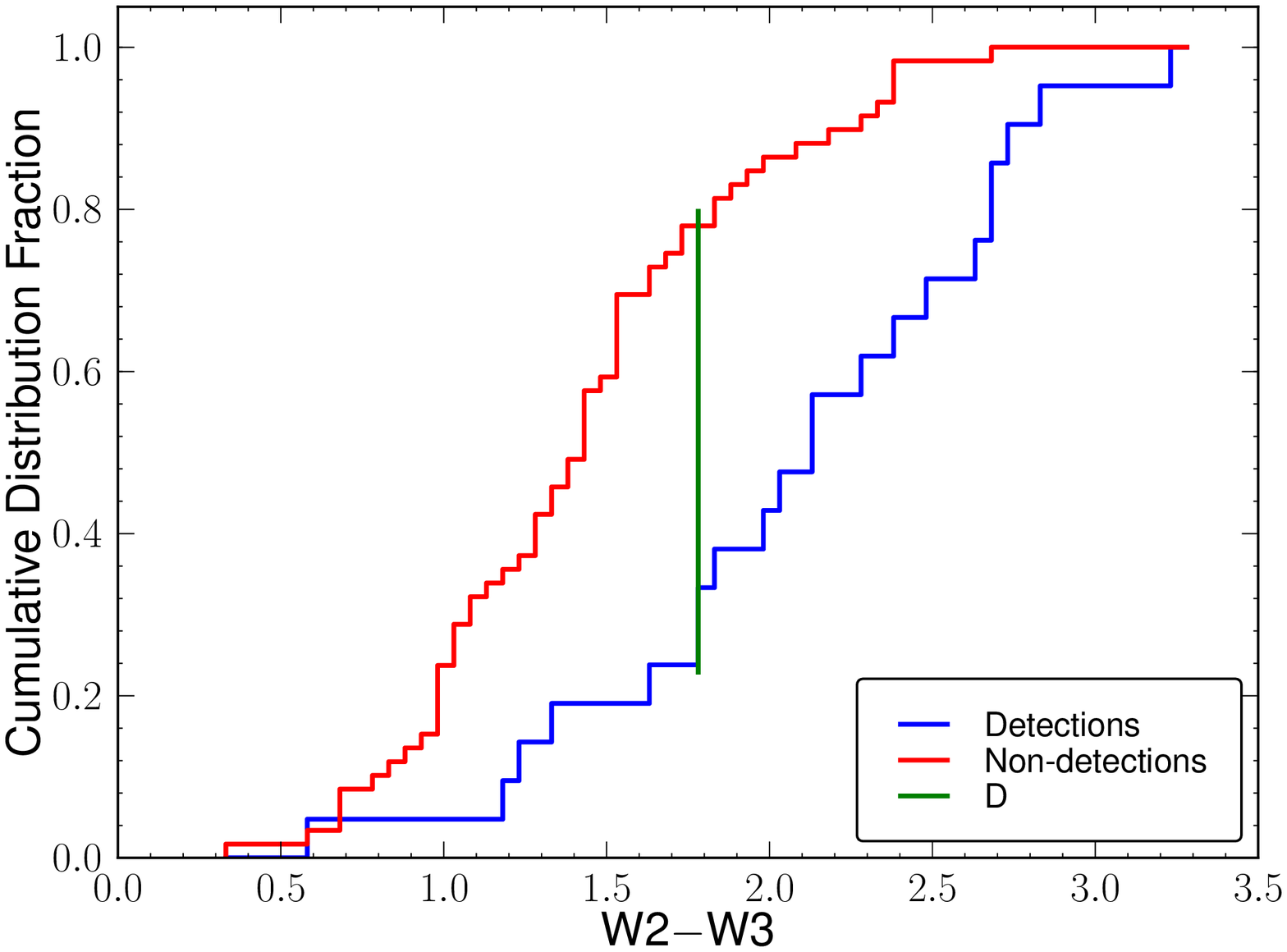}
     }  
      \caption{Left: Cumulative Distribution Fraction (CDF) of W2$-$W3 colours for all sources with H{\sc i} detections and non-detections; Right: CDF of W2$-$W3 colours for LERGs with H{\sc i} detections and non-detections. Maximum difference (D) in distributions is  shown with green vertical line.} 
     \label{fig3}
  \end{figure*}  

The  \cite{2015A&A...575A..44G} sample is dominated by LERGs (80/91) as it is a low radio luminosity/redshift sample. Therefore we largely highlight the interesting results found for LERGs; the trends for HERGs have large statistical uncertainties due to their small numbers. For example, the detection rate of H{\sc i} absorption in the HERGs is 45.5$\pm$20.3\% (5/11) compared to 26.3$\pm$5.7\% (21/80) for the LERGs. Although there is a suggestion of a higher detection rate for HERGs, it has high statistical errors and it would be useful to investigate this further using a larger number of HERGs observed to a similar sensitivity limit.

Several earlier studies have indicated a higher detection rate for compact radio sources such as compact steep-spectrum and giga-Hertz peaked spectrum sources compared with the more extended ones \citep{2003A&A...404..871P, 2006MNRAS.373..972G, 2013MNRAS.429.2380C}. Although detailed structural information at sub-arcsec level are generally not available for these sources, we examined the detection rate using the structural classification of \cite{2015A&A...575A..44G}. They classified these sources as compact and extended using the NVSS major to minor axes ratio vs the FIRST peak to integrated flux density ratios, followed by visual inspections. 
For the sources classified as either HERGs or LERGs, we find 41.3$\pm$ 9.5\% (19/46) H{\sc i} absorption detection  rate for compact ones  as compared to 15.6$\pm$5.9\% (7/45) for extended sources. In LERGs, we find  an H{\sc i} detection  rate of 38.1$\pm$9.5\% (16/42) for compact LERGs  as compared to 13.2$\pm$5.9\% (5/38) for extended LERGs.

\textit{WISE} colours are being widely used for various galaxy studies such as population studies, separating normal galaxies
from AGN, heating of dust and specific star formation rates \citep{2013AJ....145...55Y}. \cite{2014MNRAS.438..796S} noted that in their sample 92 \% of the \textit{WISE} \lq early-type' galaxies with W2$-$W3 $<$ 2 are LERGs while the \textit{WISE} \lq late type' with W2$-$W3 $\geq$ 2 are a mixture of both LERGs and HERGs. \cite{2010AJ....140.1868W} have suggested the distribution of \textit{WISE} colours for different kinds of astronomical objects in a W1$-$W2 vs W2$-$W3 diagram, and \cite{2014MNRAS.442.3361N} have modelled these in terms of dust with different temperatures and radial profiles. All colours in this paper are in Vega mag.

We checked the infrared data from \textit{WISE} survey for our sample of sources. The distributions of W2$-$W3 colours for sources with H{\sc i} absorption detection and non-detection, and locations of the different categories of objects in the W1$-$W2 vs W2$-$W3 plot (see \citealt{2010AJ....140.1868W}) are shown in Figs.~\ref{fig2} and ~\ref{fig3}. 
As we can see from these figures, there is a clear dichotomy in the distributions of W2$-$W3 colour for sources with H{\sc i} detection and non-detection. A Kolomogorov-Smirnov(KS) test rejects the null hypothesis that two samples are similar in their W2$-$W3 distribution with p-value $=$ 7.35$\times$10$^{-6}$ and D (maximum difference)$=$ 0.52 at W2$-$W3 $=$ 1.8. A similar exercise for sources with H{\sc i} detections and non-detections for only LERGs gives p $=$ 0.00011 and D$=$0.54 at W2$-$W3 $=$ 1.8, clearly indicating that these are distinct in their W2$-$W3 properties.

Of the \textit{WISE} \lq early-type' galaxies with W2$-$W3 $<$ 2, almost all (96.7$\pm$12.6 \%; 59/61) are LERGs, while amongst those with W2$-$W3 $>$ 2, 70$\pm$15.3 \% (21/30) are LERGs and 30$\pm$10 \% (9/30) are HERGs. Given the statistical uncertainties, these are consistent with the  findings of \cite{2014MNRAS.438..796S}. This dichotomy is also clearly seen when we consider the HERGs where 81.8$\pm$27.3 \% (9/11) have W2$-$W3 $>$ 2 while most of the LERGs (73.8$\pm$9.6 \%, 59/80) have W2$-$W3 $<$ 2. Therefore in this sample, the HERGs are almost entirely \emph{WISE} `late-type' (although all of them are optically early type galaxies) belonging to the region populated by LIRGs or luminous infrared galaxies \citep{2010AJ....140.1868W}, while a majority of the LERGs are early-type consisting of ellipticals.
  
Considering the \textit{WISE} `late-type' galaxies with W2$-$W3 $>$2, independent of radio structure, the H{\sc i} detection rates are similar for LERGs and HERGs, the numbers being 61.9$\pm$17.2 \% (13/21) and 55.6$\pm$24.8\% (5/9) respectively. For the WISE `early-type' galaxies with W2$-$W3 $<$ 2, only 8 of the 59 LERGs (13.6$\pm$4.8 \%) and none of the 2 HERGs have an H{\sc i} detection. The nature of the host galaxies as reflected in the \textit{WISE} colours appear to play a significant role in the detection of H{\sc i} absorption for both LERGs and HERGs.

Now let us consider the radio structure of the sources along with the \textit{WISE} colours. Of the 91 sources classified as either LERGs or HERGs, 45 are extended and 46 are compact. Of 45 extended sources, 35 have \textit{WISE} colours W2$-$W3 $<$ 2 with an H{\sc i} detection rate of 11.4$\pm$5.7\% (4/35), while the remaining 10 with W2$-$W3 $>$ 2 have a detection rate of 30$\pm$17.3 \% (3/10). Considering the 46 compact radio sources, the 26 with W2$-$W3 $<$ 2 have a detection rate of 15.4$\pm$7.7 \% (4/26), compared with an H{\sc i} detection rate of 75$\pm$19.4 \% (15/20) for those with W2$-$W3 $>$ 2.
\begin{figure}
    \centering
    \hbox{
     \includegraphics[scale=0.44]{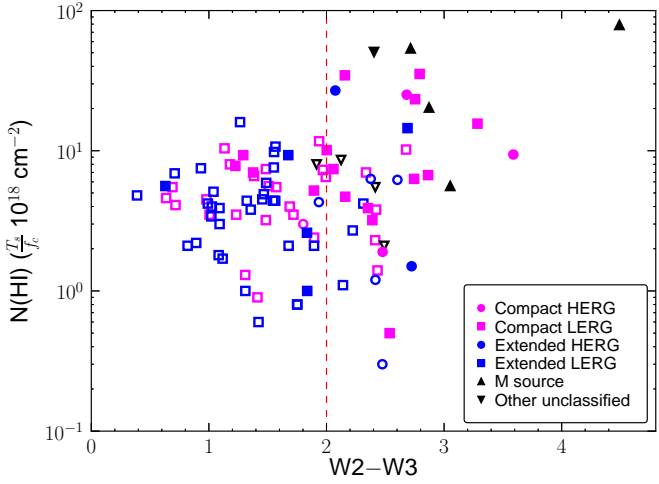}
     }
    \caption{
     Column density vs. W2$-$W3 colours for H{\sc i} absorption detections (filled symbols) and non-detections (empty symbols).}
     \label{fig4}
  \end{figure}  
 \section{Discussion} 
Our study shows a high detection rate of H{\sc i}  absorption in the \emph{WISE} `late-type' galaxies with W2$-$W3 $>$ 2 and a compact radio structure. The high H{\sc i} detection rate for the radio sources with W2$-$W3 $>$ 2 indicates a close relationship between the presence of H{\sc i} gas and warm heated dust due to either star burst or AGN. Almost all the HERGs in our sample have W2$-$W3 $>$ 2 and this can be related to mid-IR excess due to star-burst and heating of dusty torus by central AGN \citep{2016ApJ...818...65P}. We also find that the H{\sc i} absorption detection rate for LERGs  with W2$-$W3 $>$ 2 and compact radio structure is high (12/17; 70.6$\pm$20.4 \%).
Some of these LERGs have column densities comparable to HERGs and merger sources (Fig.~\ref{fig4}). This implies there is also a sub population of LERGs which is gas and dust rich, although majority of these are gas poor, as reflected in the overall detection rate. 

\begin{figure}
    \centering       
     \includegraphics[scale=0.44]{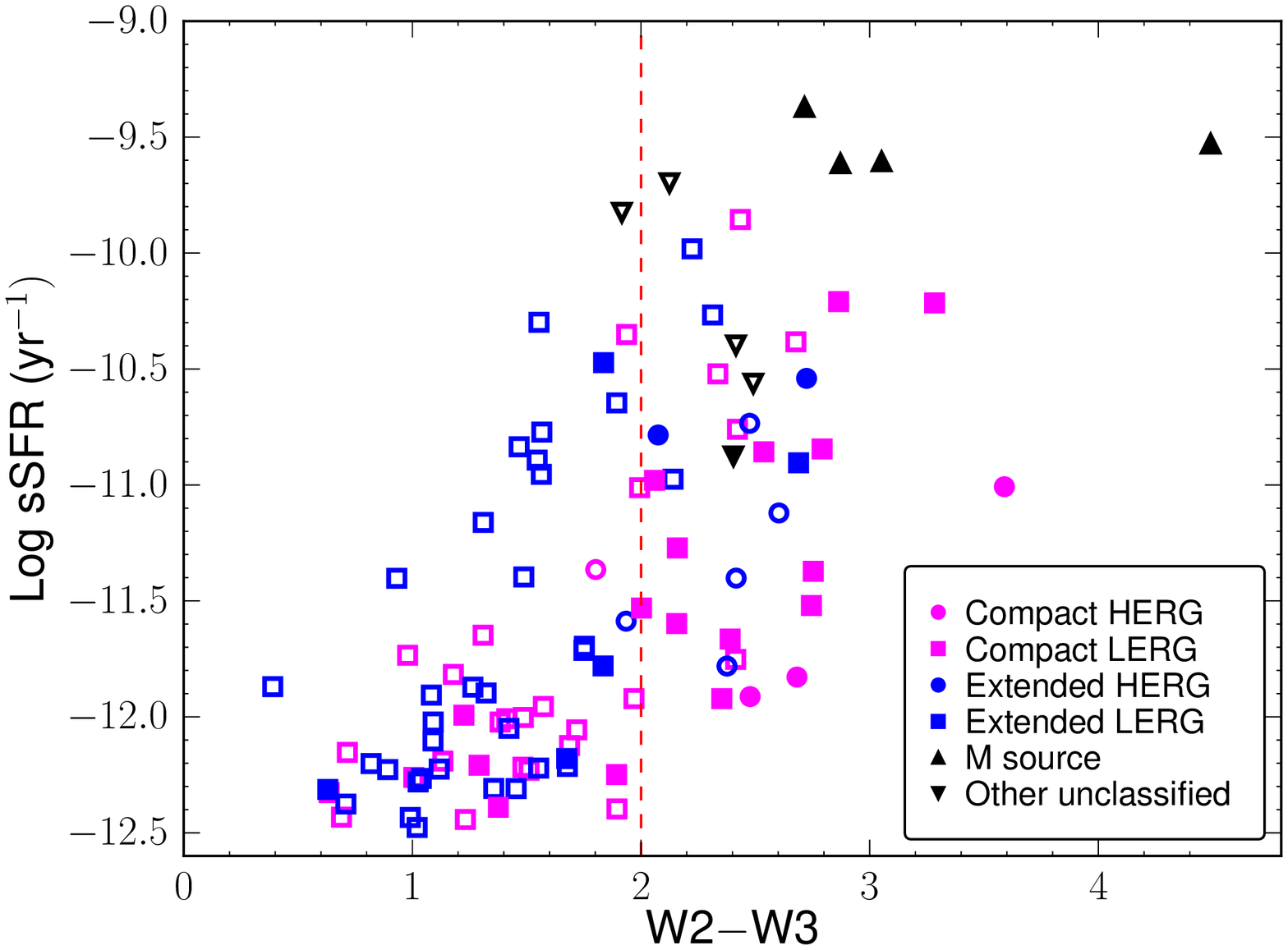}
     \includegraphics[scale=0.44]{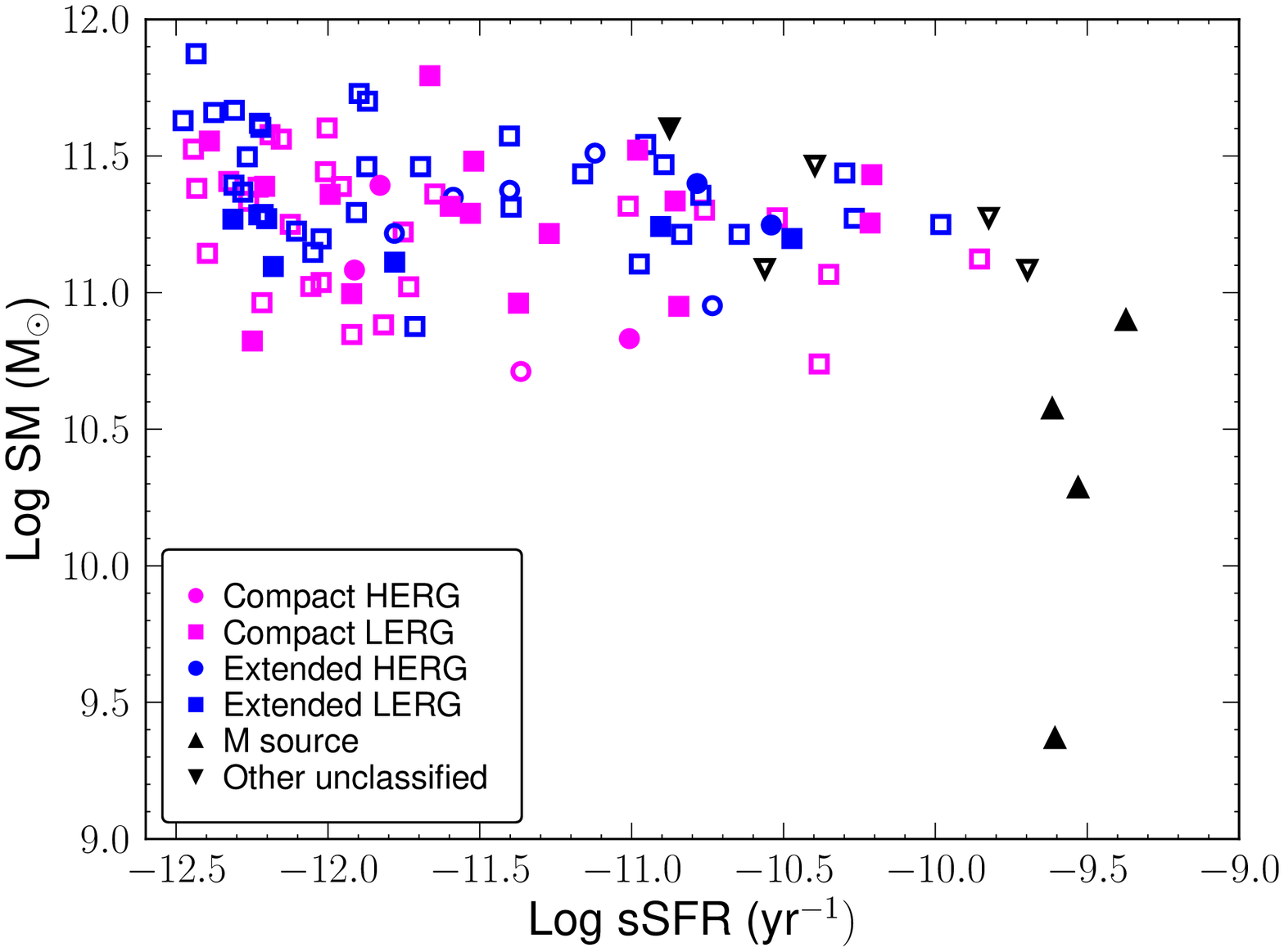} 
    \caption{
   Top: Log sSFR vs.  W2$-$W3,
   Bottom: Log Total stellar mass vs. Log sSFR, for absorption detections (filled symbols) and non-detections (empty symbols).}
     \label{fig5}
  \end{figure}      

\begin{figure*}
    \centering
    \hbox{
      \includegraphics[scale=0.44]{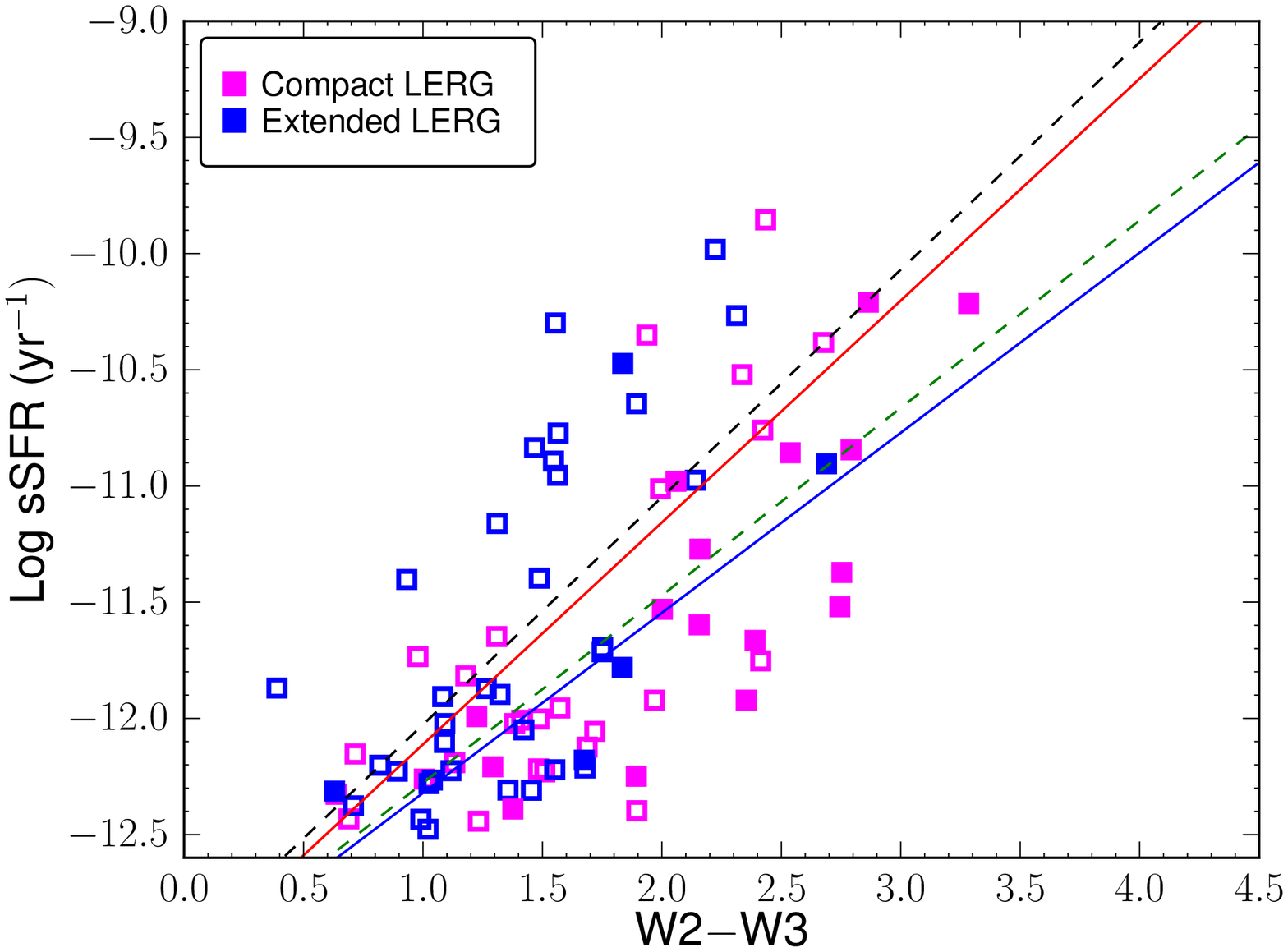}
      \includegraphics[scale=0.44]{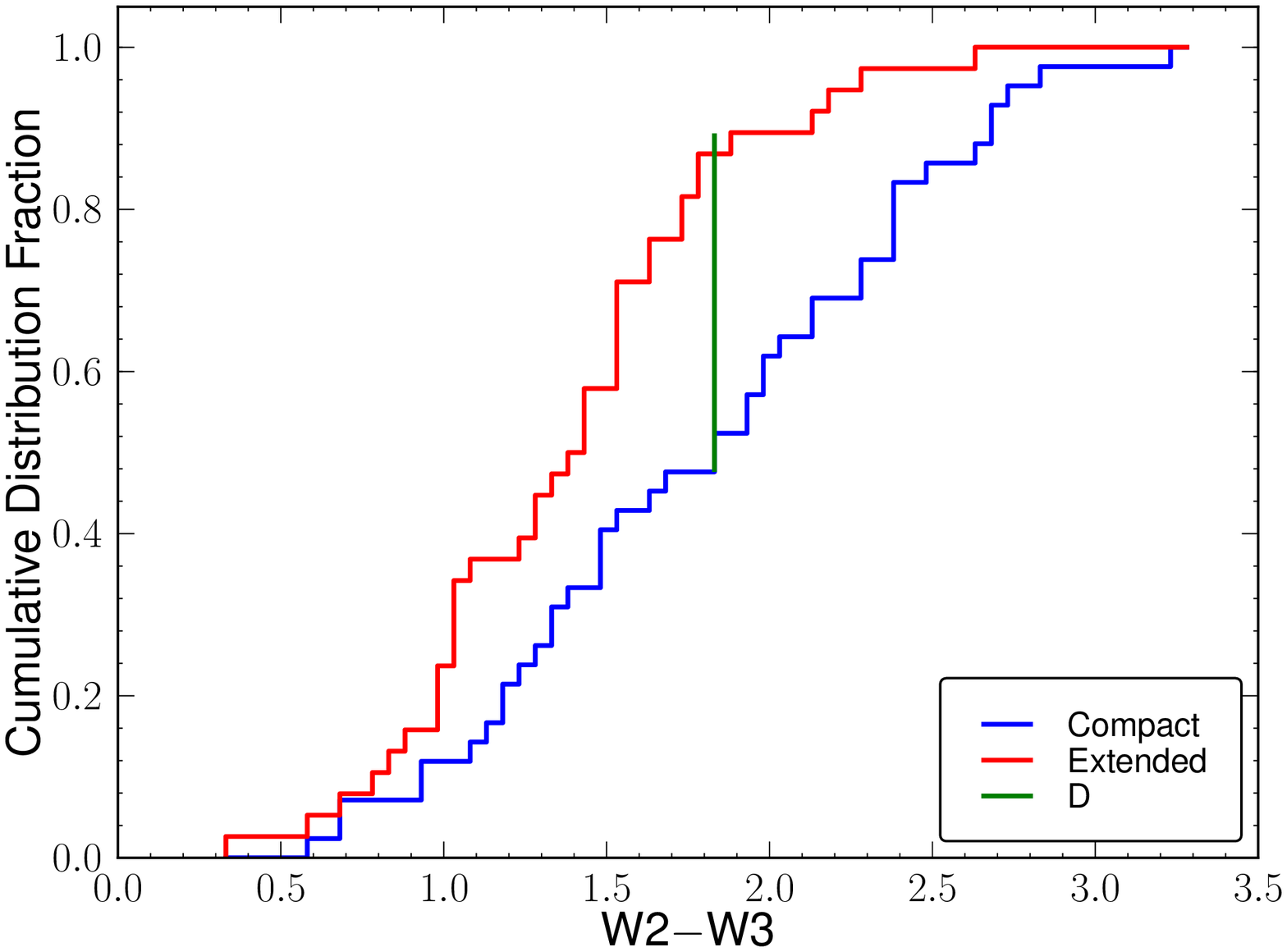}       
      }                       
    \caption{
   Left: Log sSFR vs. W2$-$W3 for LERGs (H{\sc i} detections with filled symbols and non-detections with empty symbols) in our sample with linear fits for detections (Blue solid line), compact LERGs (green dashed line), non-detections (red solid line), extended LERGs (black dashed line); 
   Right: CDF of W2$-$W3 for compact and extended LERGs where maximum difference (D) in distributions is  shown with green vertical line.
    }
     \label{fig6}
  \end{figure*}   
LERGs are believed to lack central dusty torus \citep{2006ApJ...647..161O, 2014ARA&A..52..589H} which 
indicates that in the case of gas-rich LERGs it is either individual clouds from mergers or interstellar medium playing the role of absorbers. To further investigate  these aspects we examine their stellar mass (SM), star formation rate (SFR), specific star formation rate (sSFR), and see how these relate with \emph{WISE} colours and detection rates of our sample for different categories of objects. For this we used online available information for our sample, from MPA-JHU database \footnote{\url{http://wwwmpa.mpa-garching.mpg.de/SDSS/DR7/Data/stellarmass.html}}$^{,}$\footnote{\url{http://wwwmpa.mpa-garching.mpg.de/SDSS/DR7/sfrs.html}}.  Stellar masses for these sources in this database have been derived using fits to the photometry and estimates are  similar to those of \cite{2003MNRAS.341...33K} who used spectral line indices and features like the 4000{\AA}  break. The star formation rate and specific star formation rate have been derived  using the methods of \cite{2004MNRAS.351.1151B} with corrections for dust attenuation. Also, aperture corrections were done using methodology of \cite{2007ApJS..173..267S}. Fig.~\ref{fig5} shows the variation of Log sSFR with \emph{WISE} colour for different categories of objects. In our sample, \emph{WISE} `late type' LERGs as well as HERGs are of similar stellar masses and specific star formation rates which is evident from the Fig.~\ref{fig5}. While  we have higher specific star formation rate for the objects with W2$-$W3 $>$ 2 spanning a range of  about 2 orders of magnitude with Log sSFR upto a few times $-$9 yr$^{-1}$ (Fig.~\ref{fig5}), a significant number of sources with \emph{WISE} colour W2$-$W3 $<$ 2 have Log sSFR $\lesssim$ $-$12 yr$^{-1}$.  However there are also some objects with W2$-$W3 $<$ 2 but Log sSFR $>$ $-$12 yr$^{-1}$. The median value of Log sSFR for all sources with \emph{WISE} colours W2$-$W3 $<$ 2 is $-$12.02 yr$^{-1}$ while for those with W2$-$W3 $>$ 2 the median value of Log sSFR = $-$10.85 yr$^{-1}$. Considering only LERGs, the median values of Log sSFR for those with W2$-$W3 $<$ 2 and W2$-$W3 $>$ 2 are very similar to those for the entire sample. Higher specific star formation rate towards sources with \emph{WISE} `late' type colours suggests that gas and dust rich ISM could be one of the factors affecting the detection rate in the favour of compact LERGs/HERGs. This is consistent with findings of \cite{2001MNRAS.323..331M}, who suggested that in some FR\,I type sources it is the star-forming gas in a rich ISM could play the role of absorbing gas. Another important factor affecting the detection rate is the fraction of the background source covered by the absorber given by the covering factor $f_\mathrm{c}$, which  inversely depends on size of the radio source and affects our estimate of  the integral optical depth \citep{2013MNRAS.431.3408C}. Moreover due to the compactness, compact radio sources trace sightlines towards central denser regions of radial density profiles of H{\sc i} disc \citep{2003A&A...404..871P}. This could be the other reason for the low detection rate in extended radio sources. Available VLBI images for some of the compact LERG radio sources with H{\sc i} absorption detection suggest that these  have sizes of parsec scale. However, we lack complete information on all sources. In HERGs, in addition to clouds in the interstellar medium, a dusty torus in an AGN could give rise to HI absorption if the compact radio source is of a similar or smaller scale than the torus.
\begin{figure*}
    \centering  
    \hbox{   
      \includegraphics[scale=0.44]{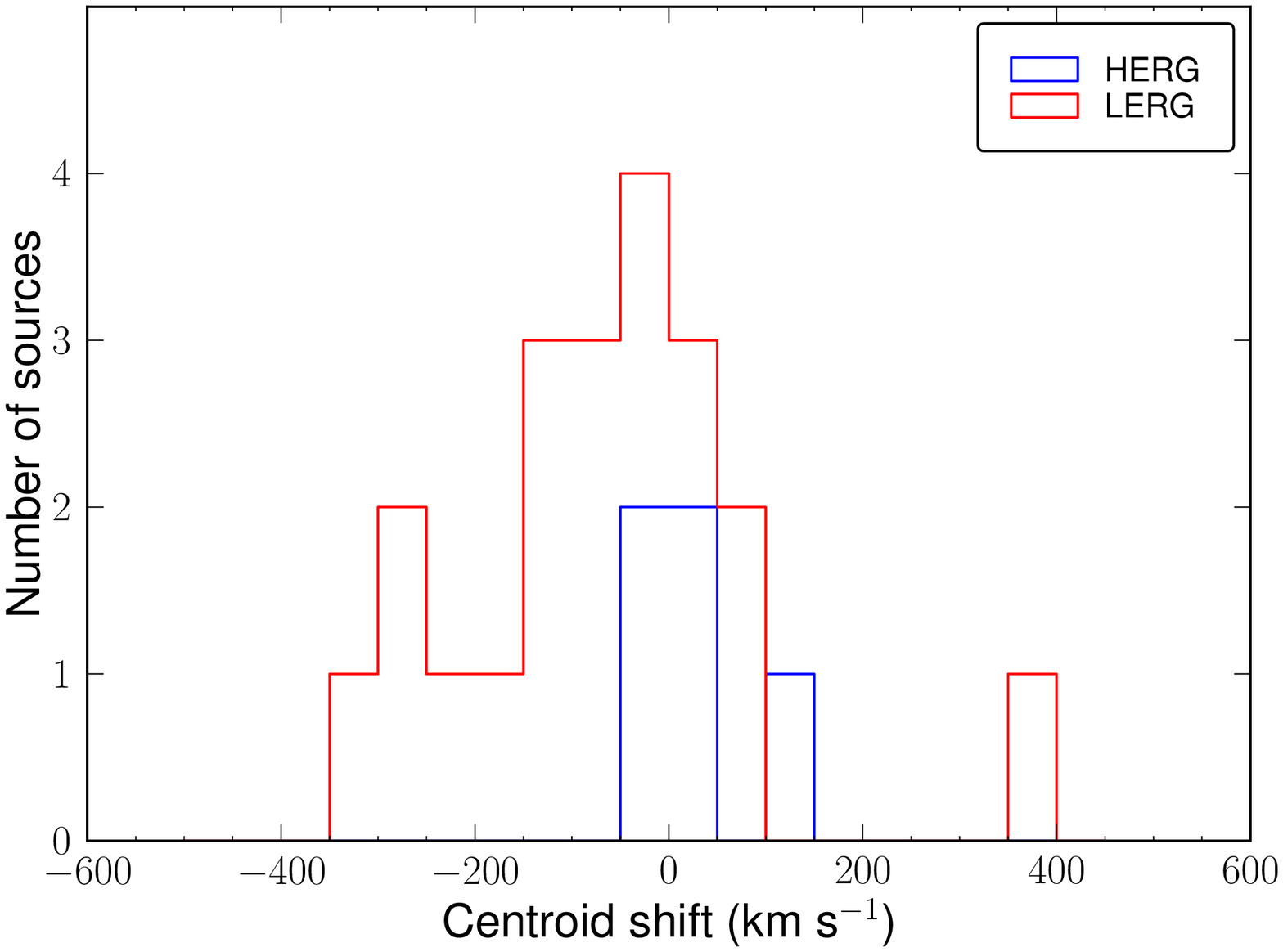}
       \includegraphics[scale=0.44]{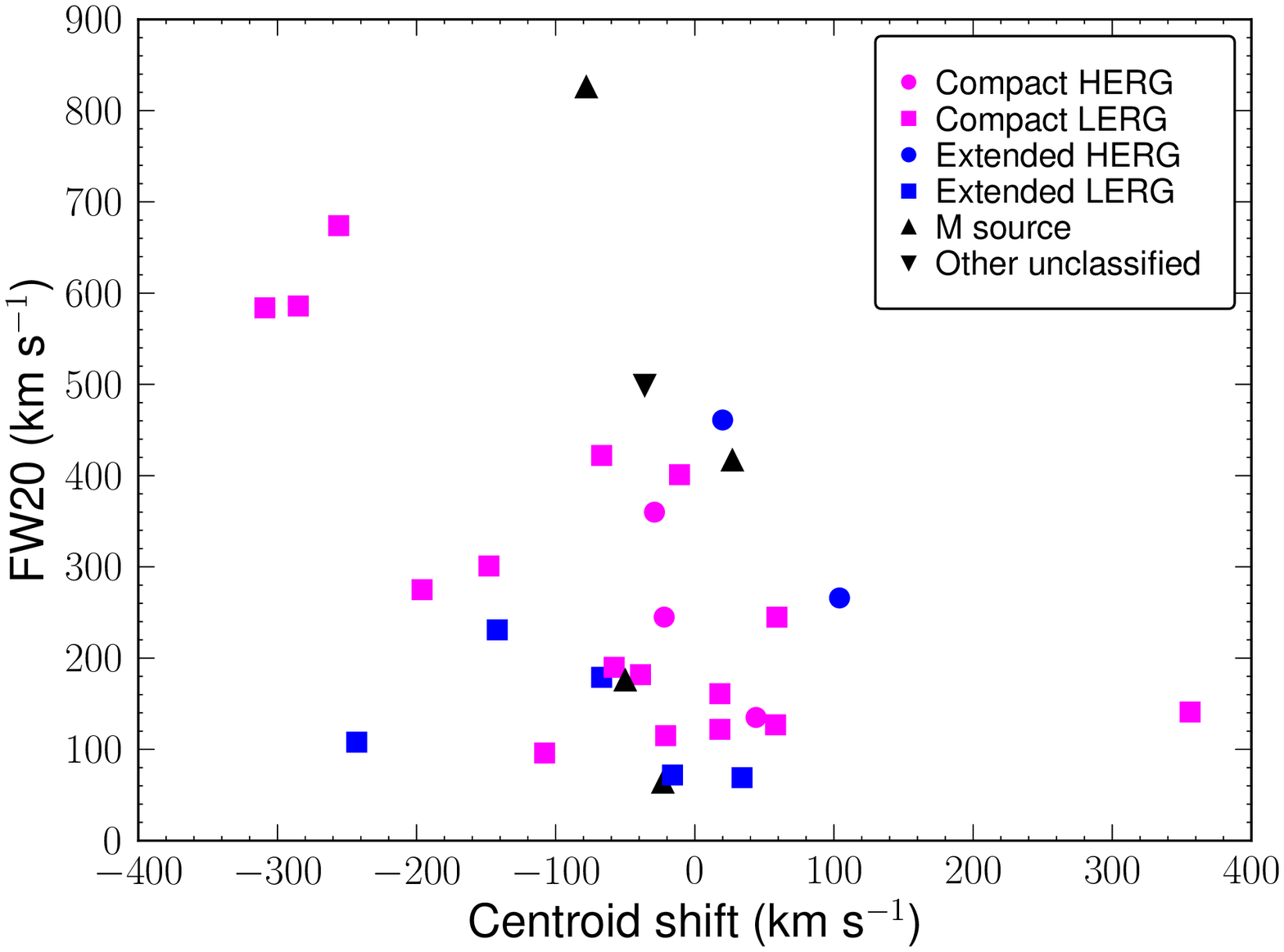}
        }
       \hbox{
      \includegraphics[scale=0.44]{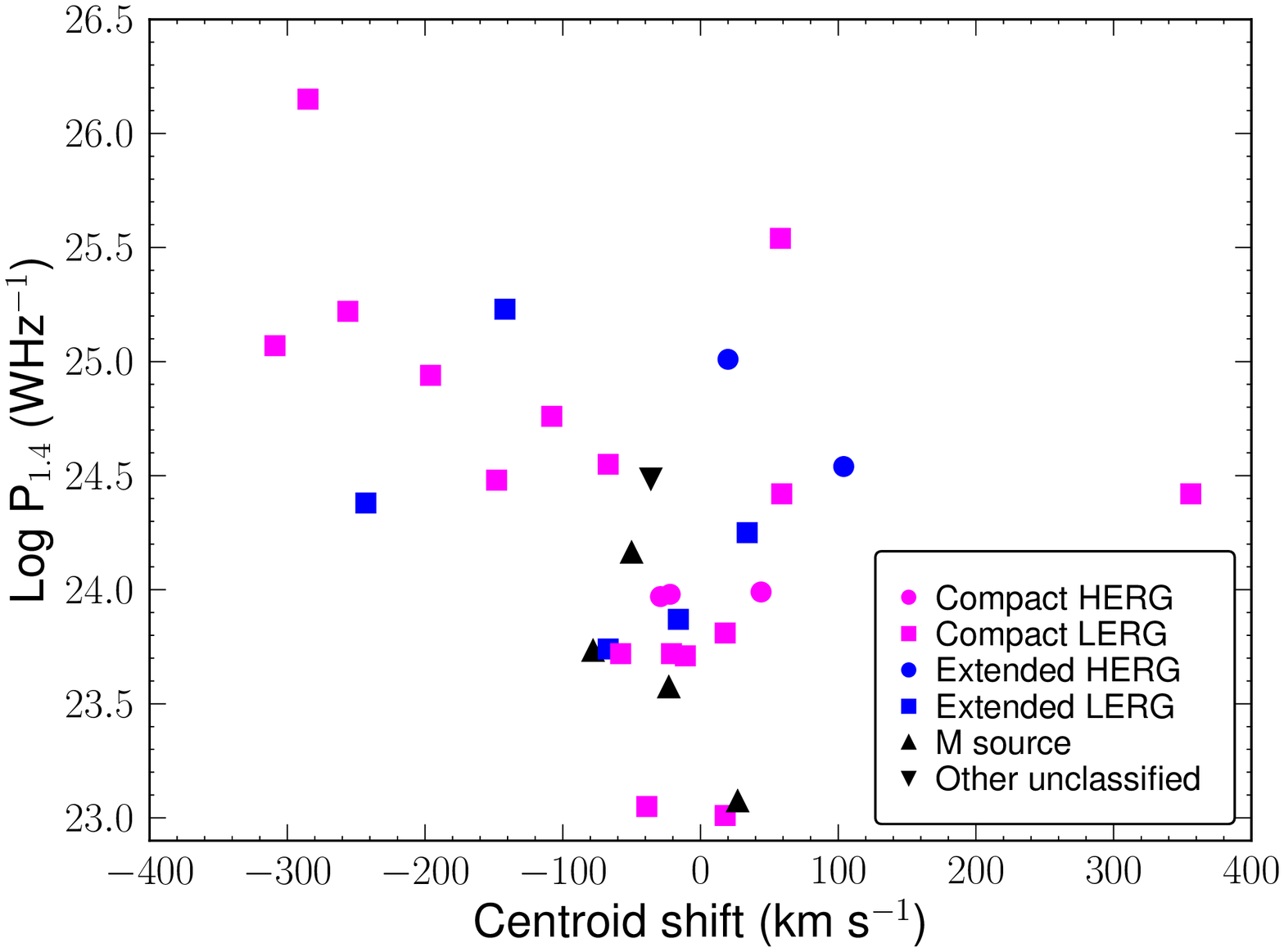}
      \includegraphics[scale=0.44]{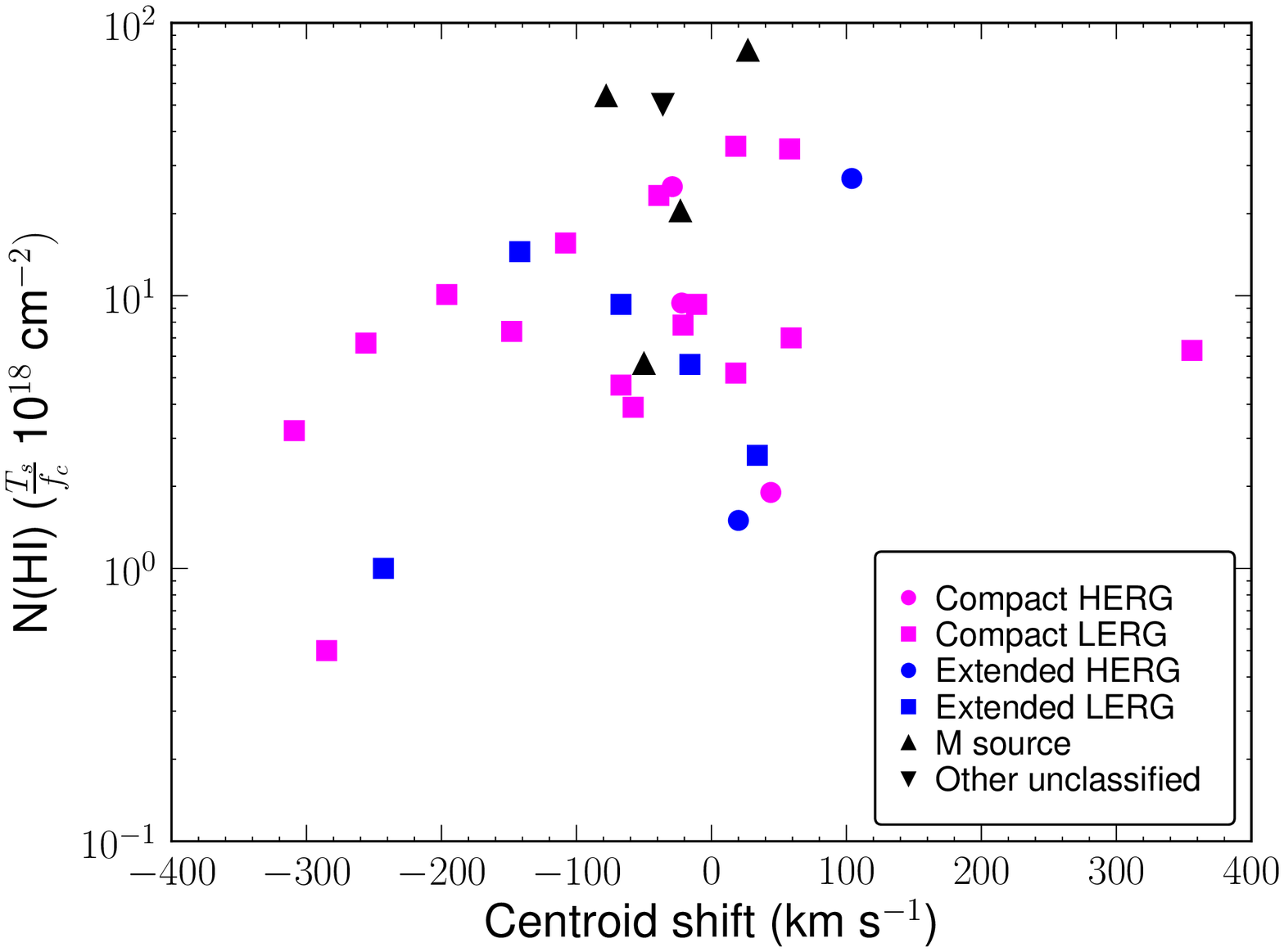}
      }
\caption{
    Top left: Distribution of centroid shift for LERGs and HERGs; Top right: FW20 (Full width at 20 percent of peak) vs. centroid shift w.r.t optical systemic velocity; 
    Bottom left: Radio luminosity at 1.4 GHz vs. centroid shift; Bottom right: Column density vs. centroid shift.}
     \label{fig7}
 \end{figure*}     

Since H{\sc i} absorption detections as well as star formation rates are higher for  
\emph{WISE}  `late-type' galaxies, it is relevant to enquire about the possible origins
of the gas or fuel. It could be the result of a normal  process of galaxy evolution 
in which star-formation  started in a dusty gas-rich environment and decreased at later stages due to scarcity of fuel, combined with possible feedback from a radio source. 
In addition, star-formation or AGN activity may get a boost from dust rich mergers or 
interactions. In the literature, suggested fuelling mechanisms vary from major mergers 
or secular processes for HERGs to accretion of halo gas, stellar remnants, winds from
evolved stars or minor mergers for LERGs \citep{2015MNRAS.451L..35E}. 
Also LERGs are reported with suppressed star-formation in the literature \citep{2015MNRAS.452.3776G, 2016ApJ...818...65P, 2016MNRAS.458L..34E}. Although this may be true for a majority of the LERGs in our sample, it doesn't explain 
the higher H{\sc i} gas detection rate and 
higher rate of star formation in \emph{WISE} `late-type' LERGs. There have been suggestions for higher rates of mergers for \emph{WISE} IR-selected AGNs with W1$-$W2 $>$ 0.8 \citep{2014MNRAS.441.1297S, 2016MNRAS.458L..34E}. However, as can be seen from Fig.~\ref{fig2}, our sample hardly has any AGN with W1$-$W2 $>$ 0.8. Also the optical host galaxies don't suggest any significant morphological differences, except a few cases of disturbed morphology or evidence of interaction with a companion galaxy e.g. J105327.2+205836. To understand these aspects and clarify any possible evolutionary scenario, we fit a first order linear function in Log sSFR vs. \emph{WISE} colour for LERG detections, non-detections, and also for compact and extended LERGs separately. Using Kendall-tau test, we find there exists a significant correlation ($\tau =$ 0.47, p $=$ 4.37$\times$10$^{-10}$) between Log sSFR and \emph{WISE} W2$-$W3 colour (Fig.~\ref{fig6}). Left panel of Fig.~\ref{fig6} also shows that compact radio sources (slope:0.81$\pm$0.12) and the detections (slope:0.78$\pm$0.16) have similar flatter slopes while extended sources (slope:0.98$\pm$0.19) and the non-detections (slope:0.96$\pm$0.14) have similar and somewhat steeper values but differences are only about 1-sigma. This correlation is consistent with the findings of \cite{2012ApJ...748...80D}, with our sample of sources lying in the region of weak AGN where sSFR is lower than
the star-forming galaxies of  \cite{2012ApJ...748...80D} by about an order of magnitude.

 Fig.~\ref{fig3} (right panel) shows that $\sim$80 \% of LERGs with no H{\sc i} detection have W2$-$W3 $<$ 1.8 which imply a significant old stellar population in their host galaxies. Also a high fraction ($\gtrsim$80 \%) of extended LERGs have 
W2$-$W3 $<$ 1.8 (right panel of Fig.~\ref{fig6}) again suggesting a significant old stellar population in their 
host galaxies compared with the compact sources. We further explored this using KS-test for W2$-$W3 and Log sSFR properties of compact and extended LERGs.  While the KS-test shows that the two samples have similar Log sSFR distributions with p value $=$ 0.94 and D $=$ 0.11, the two samples are distinct in their W2$-$W3 values with p $=$ 0.0027 (3-sigma significance) and D $=$ 0.39. A KS test also demonstrates that H{\sc i} detections and non-detections amongst compact LERGs have more significant differences in \emph{WISE}  colour 
distributions (D $=$ 0.56, p $=$ 0.0023) as compared with Log sSFR (D $=$ 0.42, p $=$ 0.044). For extended LERGs, where we have very 
few detections, these differences are of low significance in both WISE colour (D $=$ 0.58, p $=$ 0.059) and Log sSFR (D $=$ 0.19, p $=$ 0.99). Hence this difference in \emph{WISE} colours of two samples of different structures imply depletion of gas for the extended sources 
which appears to be playing an important role in H{\sc i} detection rates.

While this may indicate depletion of gas for the extended sources, we also explore any possible role of feedback from radio jets by examining the H{\sc i} absorption profiles of these sources. H{\sc i} absorption studies have shown that highly-blueshifted gas due to jet-cloud interactions often have low optical depth which may escape detection. For this discussion we have therefore used the FW20 (Full width at 20 percent of the peak) values and the centroid shift of the  profiles w.r.t optical systemic velocity provided by \cite{2015A&A...575A..44G}. While HERGs have profiles with velocity shifts $\lesssim$100 km s$^{-1}$ (upper panels Fig.~\ref{fig7}) with respect to the optical redshift, LERGs exhibit velocity shifts over a large range extending to blueshifts of about $-$310 km s$^{-1}$, with a median value of $-$58 km s$^{-1}$. There is also a  profile with large red-shift $\sim$356 km s$^{-1}$ indicating possible infall of  clouds  towards the host galaxy. The most blueshifted features  also tend to have the largest linewidths (FW20 $\gtrsim$500 km s$^{-1}$) (upper right panel of Fig.~\ref{fig7}). \cite{2015A&A...575A..44G} had shown that these highly blue-shifted (centroid shift w.r.t. optical velocity $< -$100 km s$^{-1}$) features are associated with radio sources of higher luminosity at 1.4 GHz. In the lower left panel of Fig.~\ref{fig7}, we have 
plotted our different categories of objects in a luminosity at 1.4 GHz vs. centroid shift plot. A Kendall-Tau test for Log radio luminosity vs. absolute centroid shift of LERGs gives $\tau =$ 0.51 and p $=$ 0.0013 ($>$ 3-sigma significance), indicating a significant relation as suggested earlier by \cite{2015A&A...575A..44G}. In addition, if we consider all compact sources classified as either LERGs or HERGs, a Kendall-tau test for Log radio luminosity vs. absolute centroid shift gives $\tau=$0.62 and p $=$ 0.0002 ($>$ 3-sigma significance), which is also significant. The lower right panel of Fig.~\ref{fig7} shows that the  gas with higher column densities ($>$10$^{19}$ $\frac{T_s}{f_c}$ cm$^{-2}$) have velocity shifts usually within $\pm$100 km s$^{-1}$, while the blue-shifted  LERGs show a weak tendency to have smaller values of column densities. Overall, these findings suggest that LERGs with lower radio luminosities trace the gas with less centroid shifts and some of these have high column-densities, while LERGs with relatively higher radio luminosities show evidence of outflows interacting with the ISM which may affect the star-formation rate and the \emph{WISE} colours as the sources evolve.
\section{Conclusions}
We summarise our conclusions in this section.
\begin{enumerate}
\item Earlier studies of H{\sc i} absorption had reported that compact radio sources, namely the compact steep-spectrum and giga-Hertz peaked spectrum sources, exhibited the highest detection rates of up to $\sim$45 per cent. We have shown that there is significant difference in distributions of W2$-$W3 colour for H{\sc i} absorption detections and non-detections. Considering the galaxies with \emph{WISE} infrared colour W2$-$W3 $>$ 2, which is typical of gas-rich systems, along with a compact radio structure leads to high detection rates of over 70 per cent.

\item Although majority of LERGs have low H{\sc i} detection rate, the compact LERGs with bright \emph{WISE} colours W2$-$W3 $>$ 2 also have high H{\sc i} detection rates of over 70 per cent, which indicates a gas and dust rich ISM.  

\item  The distributions of W2$-$W3 for compact and extended LERGs are significantly different, with the extended LERGs having  lower values, the maximum difference in their cumulative distributions occurring at W2$-$W3 $=$ 1.8. The W2$-$W3 colours appear to be playing an important role in determining detection of H{\sc i} in absorption rather than the specific star-formation rate.

\item Overall a lower rate of H{\sc i} detection in LERGs is consistent with a scenario of suppressed star formation rate suggested in the literature, perhaps due to feedback from radio source. It is likely that the LERGs are following a secular evolutionary process and some of these started their evolution with higher gas/dust content and may have also undergone a minor merger at some stages of their evolution, but feedback from the radio source appears to be playing a role in affecting and suppressing star formation at later stages.
  
\end{enumerate}
\section*{Acknowledgements}
The authors thank an anonymous referee for many valuable comments and suggestions which have helped improve the paper significantly.This work of YC was sponsored by the Chinese Academy of Sciences Visiting Fellowship for Researchers from Developing Countries, Grant No. 2013FFJB0009”. YC also acknowledges support from National Natural Science Foundation of China (NSFC) Grant No. 11550110181 and the China Ministry of Science and Technology under State Key Development Program for Basic Research (973 program) No. 2012CB821802.
 This publication makes use of data products from the \emph{Wide-field Infrared Survey Explorer}, which is a joint project of the University of California, Los Angeles, and the Jet Propulsion Laboratory/California Institute of Technology, funded by the National Aeronautics and Space Administration. This research has made use of NASA's Astrophysics Data System. This research has made use of the NASA/IPAC Extragalactic Database (NED) which is operated by the Jet Propulsion Laboratory, California Institute of Technology, under contract with the National Aeronautics and Space Administration. This research has made use of the VizieR catalogue access tool, CDS, Strasbourg, France. The original description of the VizieR service was published in A\&AS 143, 23.
 This work also makes use of \emph{Sloan Digital Sky Survey} (SDSS)-III . Funding for SDSS-III has been provided by the Alfred P. Sloan Foundation, the Participating Institutions, the National Science Foundation, and the U.S. Department of Energy Office of Science. The SDSS-III web site is \url{http://www.sdss3.org/}. SDSS-III is managed by the Astrophysical Research Consortium for the Participating Institutions of the SDSS-III Collaboration including the University of Arizona, the Brazilian Participation Group, Brookhaven National Laboratory, Carnegie Mellon University, University of Florida, the French Participation Group, the German Participation Group, Harvard University, the Instituto de Astrofisica de Canarias, the Michigan State/Notre Dame/JINA Participation Group, Johns Hopkins University, Lawrence Berkeley National Laboratory, Max Planck Institute for Astrophysics, Max Planck Institute for Extraterrestrial Physics, New Mexico State University, New York University, Ohio State University, Pennsylvania State University, University of Portsmouth, Princeton University, the Spanish Participation Group, University of Tokyo, University of Utah, Vanderbilt University, University of Virginia, University of Washington, and Yale University. This work has also used different Python packages e.g. numpy, scipy and matplotlib.  

\newpage
\appendix
\begin{deluxetable}{cccccccccccc}
\tablewidth{0pt}
\rotate
\tabletypesize{\scriptsize}
\label{aptable1}
\tablecaption{Characteristics of optical  lines  for 31 sources detected in H{\textsc i} absorption by Ger\'eb et al. (2015). Full table in online version.}
\startdata
\hline
(1)&(2)&(3)&(4)&(5)&(6)&(7)&(8)&(9)&(10)&(11)&(12)\\
Sq. & RA & Dec & H$_{\beta}$ Flux & O[{\sc{iii}}] Flux & O[{\sc{iii}}] EQW & O[{\sc{i}}] Flux & H$_{\alpha}$ Flux & N[{\sc{ii}}] Flux & S[{\sc{ii}}] Flux & Optical & Radio \\
&&&&&\AA&&&&&Class&Class                                                 \\
\hline
1 & 07 57 56.7 & +39 59 36 & 186.261(    10.098) & 1364.281(    20.400) & 37.224(     0.360) & 105.589(     8.243) & 803.512(    23.928) & 691.473(    16.945) & 257.189(    10.055) & HERG & C \\
2 & 08 06 01.5 & +19 06 15 & 46.529(     8.398) & 376.843(    10.831) & 11.258(     0.222) & 43.195(     6.408) & 336.209(    17.030) & 381.474(    13.136) & 136.350(     9.655) & HERG & E \\
3 & 08 09 38.9 & +34 55 37 & 91.580(    13.851) & 44.482(     8.970) & 1.059(     0.134) & 18.584(     7.407) & 114.215(    21.062) & 132.103(    15.357) & 34.271(    10.370) & LERG & E \\
4 & 08 36 37.8 & +44 01 10 & 141.161(    11.989) & 915.593(    17.147) & 23.732(     0.282) & 114.595(     9.000) & 528.403(    26.060) & 632.954(    20.318) & 254.389(    13.003) & HERG & C \\
5 & 08 43 07.1 & +45 37 43 & 6.874(     4.445) & 6.719(     4.618) & 0.908(     0.462) & 1.435(     3.411) & 28.736(    13.966) & 29.787(    10.209) & 2.635(     5.381) & LERG & C \\
6 & 09 09 37.4 & +19 28 08 & 254.722(    18.626) & 342.203(    16.477) & 3.349(     0.108) & 258.302(    17.929) & 901.681(    48.750) & 957.775(    37.700) & 593.890(    26.776) & LERG & C \\
.... & .... & .... & .... & .... & .... & .... & .... &....& .... & .... & .... \\

\hline
\enddata
\begin{flushleft}\textbf{Notes:} Column 1: Sequence no. ; Columns 2-3: co-oridinates for radio sources from Ger\'eb et al. (2015); columns 4-10: line parameters for different optical lines (with errors) from MPA-JHU group database (Brinchmann et al. 2004a); column 11: Optical classification: HERG/LERG/U(unclassified) following Best \& Heckman (2012); column 12: radio structural classification C (compact) E(extended) M (Merger/optical blue objects) from Ger\'eb et al. (2015) \\
Flux values are in units of 10$^{-17}$ ergs s$^{-1}$ cm$^{-2}$ and errors are scaled up by factors as recommended by MPA-JHU group (Brinchmann et al. 2004a). In column 6, O[{\sc iii}] equivalent width (EQW) is in Angstrom. Only absolute values for these measurements are provided in this table.\\
$^{\dagger}$ for absorption lines with equivalent width positive.
\end{flushleft}
\end{deluxetable}
\begin{deluxetable}{cccccccccccc}
\tablewidth{0pt}
\rotate
\tabletypesize{\scriptsize}
\label{aptable2}
\tablecaption{Characteristics of optical  lines  for 69 sources not detected in H{\sc i} absorption by Ger\'eb et al. (2015). Full table in online version.}
\startdata
\hline
(1)&(2)&(3)&(4)&(5)&(6)&(7)&(8)&(9)&(10)&(11)&12\\
Sq.& RA & Dec & H$_{\beta}$ Flux & O[{\sc{iii}}] Flux & O[{\sc{iii}}] EQW & O[{\sc{i}}] Flux & H$_{\alpha}$ Flux & N[{\sc{ii}}] Flux & S[{\sc{ii}}] Flux & Optical & Radio \\
&&&&&\AA&&&&&Class&Class                                                 \\
\hline
1 & 07 56 07.1 & +38 34 01 & 14.764(     5.731) & 14.764(     4.364) & 1.698(     0.395) & 9.303(     4.089) & 46.291(    11.643) & 109.312(    10.642) & 26.828(     5.360) & LERG & C \\
2 & 07 58 28.1 & +37 47 12 & 125.398(    22.360) & 278.040(    19.919) & 2.571(     0.121) & 119.835(    19.460) & 473.809(    54.911) & 922.909(    45.210) & 150.965(    24.682) & LERG & E \\
3 & 07 58 47.0 & +27 05 16 & 52.036(     6.965) & 37.139(     4.774) & 2.213(     0.209) & 21.922(     3.793) & 102.876(    10.638) & 69.520(     6.935) & 27.495(     4.739) & LERG & C \\
4 & 08 00 42.0 & +32 17 28 & 4.400(    11.869) & 11.738(     8.383) & 1.278(     0.391) & 2.331(     6.052) & 9.035(    19.401) & 19.641(    13.878) & 1.880(     9.496)$^{\dagger}$ & LERG & E \\
5 & 08 18 27.3 & +28 14 03 & 98.942(    11.180) & 7.570(     7.961) & 0.177(     0.201) & 3.206(     8.697) & 81.246(    16.620) & 24.049(    13.662) & 0.213(    10.966) & U & C \\
6 & 08 18 54.1 & +22 47 45 & 24.458(     8.422) & 52.467(     6.703) & 1.533(     0.139) & 24.411(     5.844) & 86.687(    16.420) & 142.249(    12.049) & 35.152(     7.774) & LERG & E \\
.... & .... & .... & .... & .... & .... & .... & .... &....& .... & .... & .... \\

\hline
\enddata
\begin{flushleft}\textbf{Notes:} Column 1: Sequence no. ; Columns 2-3: co-oridinates for radio sources from Ger\'eb et al. (2015); columns 4-10: line parameters for different optical lines (with errors) from MPA-JHU group database (Brinchmann et al. 2004a); column 11: Optical classification: HERG/LERG/U(unclassified) following Best \& Heckman (2012); column 12: radio structural classification C (compact) E(extended) M (Merger/optical blue objects) from Ger\'eb et al. (2015).\\
Flux values are in units of 10$^{-17}$ ergs s$^{-1}$ cm$^{-2}$ and errors are scaled up by factors as recommended by MPA-JHU group (Brinchmann et al. 2004a). In column 6, O[{\sc iii}] equivalent width (EQW) is in Angstrom. Only absolute values for these measurements are provided in this table.\\
$^{\dagger}$: absorption lines with equivalent width positive.
\end{flushleft}
\end{deluxetable}
\hspace{1cm}
\begin{deluxetable}{cccccccccc}
\tablewidth{0pt}
\rotate
\tabletypesize{\scriptsize}
\label{aptable3}
\tablecaption{\emph{WISE} magnitudes, Log stellar masses (SM), Log star formation rates (SFR) and Log specific star formation rates (sSFR) for all 100 sources. Full table in online version}

\startdata
\hline
   (1)     & (2) & (3)  & (4) & (5)  & (6) & (7 ) &  (8) & (9) \\
Sq. & ALLWISE & W1mag & W2mag & W3mag & W4mag & Log SM & Log SFR & Log sSFR \\
 &name        & &  &  &  & M$_{\odot}$& M$_{\odot}$yr$^{-1}$& yr$^{-1}$ \\
\hline
1 & J075607.17+383400.4 & 14.101(     0.043) & 13.750(     0.052) & 12.265(        .....   ) & 8.980(        .....   ) & 11.602 & -0.342 & -12.003 \\
2 & J075756.71+395936.1 & 12.832(     0.023) & 12.338(     0.024) & 8.749(     0.028) & 5.088(     0.030) & 10.831 & -0.121 & -11.008 \\
3 & J075828.11+374711.8 & 10.837(     0.021) & 10.788(     0.021) & 9.671(     0.052) & 7.764(     0.254) & 11.620 & -0.554 & -12.226 \\
4 & J075847.02+270515.7 & 12.877(     0.025) & 12.226(     0.022) & 9.549(     0.044) & 7.472(     0.166) & 10.739 & 0.381 & -10.383 \\
5 & J080041.93+321728.8 & 13.789(     0.038) & 13.514(     0.043) & 12.192(        .....   ) & 8.239(        .....   ) & 11.729 & -0.091 & -11.897 \\
6 & J080601.54+190614.6 & 12.462(     0.026) & 12.195(     0.026) & 10.120(     0.068) & 8.430(     0.443) & 11.399 & 0.662 & -10.785 \\
7 & J080938.85+345537.3 & 12.477(     0.026) & 12.176(     0.025) & 10.340(     0.092) & 8.218(     0.422) & 11.198 & 0.783 & -10.473 \\
.... & .... & .... & .... & .... & .... & .... & .... & .... \\
.... & .... & .... & .... & .... & .... & .... & .... & .... \\
\hline
\enddata
\begin{flushleft}\textbf{Notes:} Column 1: Sequence number; column 2: \emph{WISE} name of sources; columns 3-6: WISE  vega magnitudes at W1(3.4 $\mu$m), W2(4.6 $\mu$m), W3(12 $\mu$m) and W4(22 $\mu$m) with errors from \emph{WISE} catalog (Cutri et al. 2013); column 7: stellar masses (SM) from MPA-JPU database (Brinchmann et al. 2004a,2004b); column 8: star formation rates (SFR) from MPA-JPU database (Brinchmann et al. 2004a,2004b); column 9: specific star formation rates (sSFR) from MPA-JPU database (Brinchmann et al. 2004a,2004b).\\
Upper limits on magnitudes are given with errors as blank (....).
\end{flushleft}
\end{deluxetable}
\bsp	
\label{lastpage}
\end{document}